\documentclass[reprint,prb,superscriptaddress,showpacs,aps,floatfix]{revtex4-2}%
\usepackage{graphicx}
\usepackage{soul}
\usepackage{color}
\usepackage{amsmath}
\usepackage{amsfonts}
\usepackage{amssymb}
\usepackage{framed}
\usepackage{array} 
\usepackage{tabularx} 
\usepackage{float}
\usepackage{booktabs}
\usepackage[normalem]{ulem}
\providecommand{\U}[1]{\protect\rule{.1in}{.1in}}
\makeatletter
\renewcommand*{\fnum@figure}{{\normalfont\bfseries \figurename~\thefigure}}

\renewcommand*{\@caption@fignum@sep}{\textbf{ : }}
\makeatother
\usepackage{hyperref}
\hypersetup{
    colorlinks=true,
    linkcolor=blue,
    filecolor=magenta,      
    urlcolor=cyan,
}
\begin{document}

\title{ Crystal Growth and Physical Properties of Orthorhombic Kagome Lattice Magnets $R$Fe$_6$Ge$_6$ ($R$=Y, Tb, Dy)}

\author{Abhijeet Nayak}
\email{Corresponding author: anayak@nd.edu}
\affiliation{Department of Physics and Astronomy, University of Notre Dame, Notre Dame, IN 46556, USA}
\affiliation{Stavropoulos Center for Complex Quantum Matter, University of Notre Dame, Notre Dame, IN 46556, USA}

\author{Sk Jamaluddin}
\affiliation{Department of Physics and Astronomy, University of Notre Dame, Notre Dame, IN 46556, USA}
\affiliation{Stavropoulos Center for Complex Quantum Matter, University of Notre Dame, Notre Dame, IN 46556, USA}

\author{Fan Wu}
\affiliation{Department of Physics and Astronomy, University of Notre Dame, Notre Dame, IN 46556, USA}
\affiliation{Stavropoulos Center for Complex Quantum Matter, University of Notre Dame, Notre Dame, IN 46556, USA}

\author{Emily Rapp}
\affiliation{Physics, Astronomy and Materials Science Department, Missouri State University, Springfield, MO 65897, USA}

\author{Resham Babu Regmi}
\affiliation{Department of Physics and Astronomy, University of Notre Dame, Notre Dame, IN 46556, USA}
\affiliation{Stavropoulos Center for Complex Quantum Matter, University of Notre Dame, Notre Dame, IN 46556, USA}

\author{Mohamed El Gazzah}
\affiliation{Department of Physics and Astronomy, University of Notre Dame, Notre Dame, IN 46556, USA}
\affiliation{Stavropoulos Center for Complex Quantum Matter, University of Notre Dame, Notre Dame, IN 46556, USA}

\author{Bence G. M\'arkus}
\affiliation{Department of Physics and Astronomy, University of Notre Dame, Notre Dame, IN 46556, USA}
\affiliation{Stavropoulos Center for Complex Quantum Matter, University of Notre Dame, Notre Dame, IN 46556, USA}

\author{L\'aszl\'o Forr\'o}
\affiliation{Department of Physics and Astronomy, University of Notre Dame, Notre Dame, IN 46556, USA}
\affiliation{Stavropoulos Center for Complex Quantum Matter, University of Notre Dame, Notre Dame, IN 46556, USA}

\author{Madhav P. Ghimire}
\affiliation{Central Department of Physics, Tribhuvan University, Kirtipur, Kathmandu 44613,, Nepal}
\affiliation{Leibniz Institute for Solid State and Materials Research, IFW Dresden, Helmholtzstr. 20, D01069 Dresden, Germany}

\author{Allen Oliver}
\affiliation{Department of Chemistry \& Biochemistry, University of Notre Dame, Notre Dame, IN 46556, USA}

\author{Kateryna Foyevtsova}
\affiliation{Department of Physics and Astronomy, University of Notre Dame, Notre Dame, IN 46556, USA}
\affiliation{Stavropoulos Center for Complex Quantum Matter, University of Notre Dame, Notre Dame, IN 46556, USA}

\author{Igor I. Mazin}
\affiliation{Department of Physics and Astronomy, George Mason University, Fairfax, VA 22030, USA}
\affiliation{Quantum Science and Engineering Center, George Mason University, Fairfax, VA 22030, USA}

\author{Nirmal J. Ghimire}
\affiliation{Department of Physics and Astronomy, University of Notre Dame, Notre Dame, IN 46556, USA}
\affiliation{Stavropoulos Center for Complex Quantum Matter, University of Notre Dame, Notre Dame, IN 46556, USA}

\date{\today}
\begin{abstract}
Kagome magnets represent a promising class of materials that exhibit intriguing electronic and magnetic properties, and they have recently garnered significant attention. While most kagome-lattice compounds are hexagonal, we report here single-crystal growth and physical property measurements of $R$Fe$_6$Ge$_6$ ($R$ = Y, Dy, Tb) compounds, which crystallize in an orthorhombic  structure. The structure can be derived from a hexagonal prototype $R$Fe$_3$Ge$_2$ by replacing every other $R$ atom with a covalent Ge$_2$ dimer. Ordering of these dimers renders the structure orthorhombic, slightly distorts the kagome net, and makes the three Fe sites formally inequivalent. The iron and rare-earth sublattices order independently. Fe moments order above 400 K, forming ferromagnetic kagome planes stacked antiferromagnetically, while rare-earth moments order below 9 K. TbFe$_6$Ge$_6$ exhibits a single magnetic ordering transition associated with the Tb atoms, whereas DyFe$_6$Ge$_6$ shows two distinct magnetic phase transitions, strongly influenced by crystal electric field effects on the Dy$^{3+}$ ions. Density functional theory (DFT) calculations indicate that the ferromagnetic ordering of the Fe planes is driven by a high density of states at the Fermi energy. They also reveal three dramatically different structural energy scales: $R$ and Ge$_2$ form alternating 1D chains perpendicular to the kagome planes, and violating this alternation incurs a large energy cost. Aligning these chains is less costly, and achieving a two-dimensional order of anti-aligned chains requires very little energy. These compounds represent a unique class of materials, offering new opportunities to investigate the interplay between the distinct crystal lattice geometry and the underlying electronic and magnetic properties.  

\end{abstract}

\maketitle

\section{Introduction}\label{sec:1}

\begin{figure*}[htbp] 
    \centering
    \includegraphics[width=1.0\textwidth]{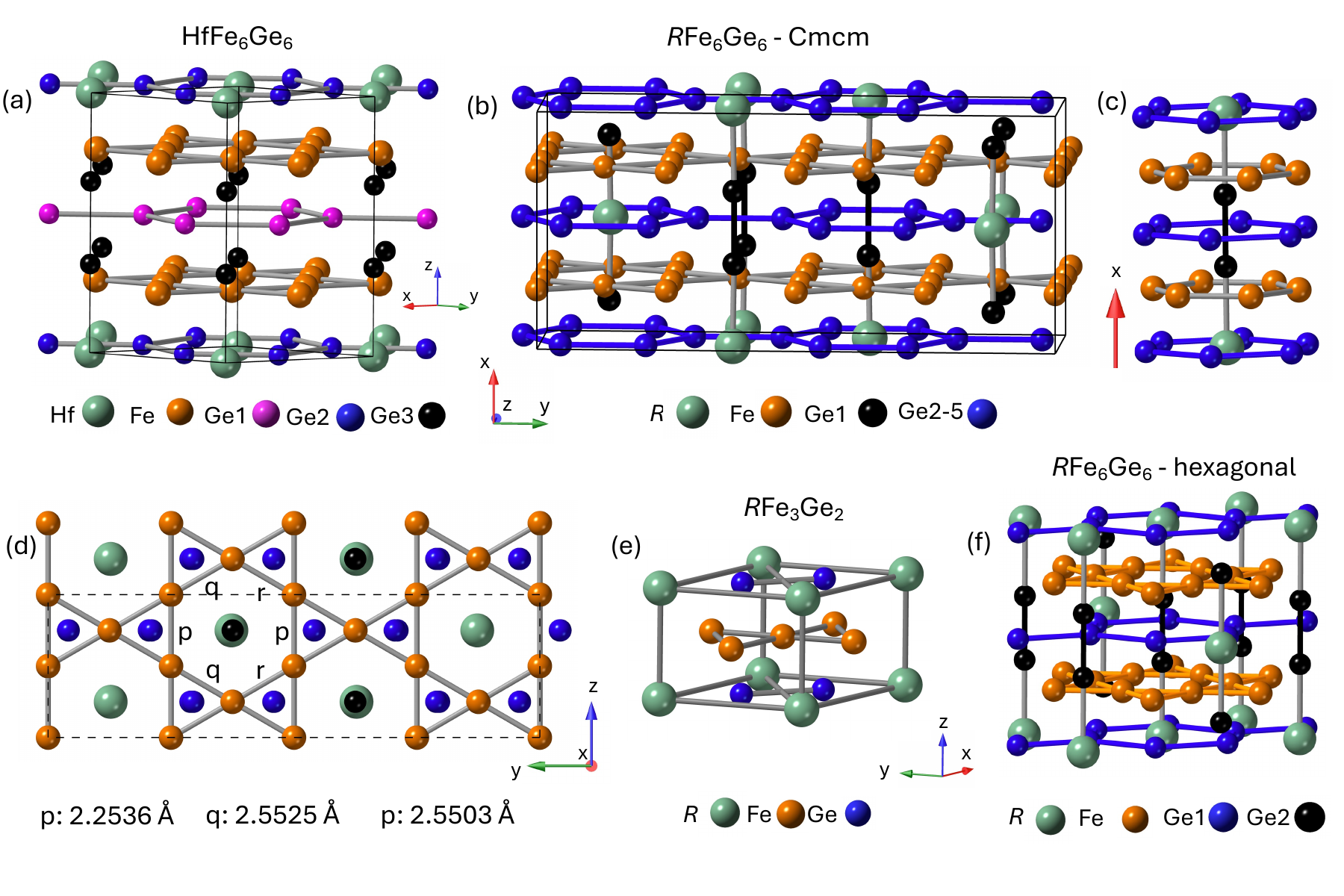}
     \caption{\small a) Sketch of HfFe$_6$Ge$_6$-type crystal structure. b) Crystal structure of $R$Fe$_6$Ge$_6$ in the Cmcm space group. There are five inequivalent Ge atoms. A different color (black) is used for Ge1 to highlight its dimerization along a-axis. Other Ge atoms (Ge2-5) are all represented by blue colored balls. c) Ge–Ge dimers forming linear chains with $R$ atoms along the $a$-axis penetrating the Ge honeycomb and Fe kagome planes. d) a-axis view of the  TbFe$_6$Ge$_6$-Cmcm structure highlighting the kagome network of Fe atoms. The bond lengths of atoms forming the kagome network slightly distorted due to the orthorhombic symmetry. The letters p, q, and r denote equivalent bond lengths within each kagome unit, as indicated in the legend. e) Sketch of a $R$Fe$_3$Ge$_2$ unit cell, a building block for the hexagonal $R$Fe$_6$Ge$_6$ structure. f) Schematic illustration of the formation of $R$Fe$_6$Ge$_6$ structure from doubling of $R$Fe$_3$Ge$_2$ (details are provided in the introduction section). The solid black lines in panels (a) and (b) and the dashed lines in panel (d) represent the unit cell of the corresponding crystal structures.}
    \label{Structure 1}
\end{figure*}

Kagome magnets have emerged as a central focus in condensed matter research due to their intriguing electronic and magnetic properties \cite{meduri2025evolution, ghimire2020topology}. Recently, it has been recognized that even subtle deviations from an ideal kagome lattice can give rise to novel and exotic physical phenomena. For instance, breathing kagome lattices host electronic and magnetic topological states \cite{bolens2019topological, hirschberger2019skyrmion, xie2025manipulation}, while twisted kagome lattice compounds exhibit electronic topological phases and spin-ice-like behavior \cite{bhandari2025tunable, zhao2024discrete, zhao2020realization}. More recently, $R$Ti$_3$Bi$_4$ compounds have drawn attention for their distinctive electronic and magnetic properties \cite{ortiz2023evolution, sakhya2024diverse, park2025spin, cheng2024spectroscopic, yang2025two, han2025discovery}. These compounds feature a distorted kagome net of Ti atoms combined with zigzag chains of $R$ atoms, an arrangement believed to produce magnetic and electronic characteristics distinct from those of other kagome-based materials.

Among the most extensively investigated kagome systems are the $RT_6X_6$ compounds, commonly referred to as $R$166, where $R$ is a rare-earth element, $T$ is a transition metal, and $X$ is Sn or Ge. These materials exhibit a remarkable variety of emergent magnetic and electronic behaviors \cite{ghimire2020competing, wang2021field, roychowdhury2022large, li2022manipulation, kitaori2021emergent, bhandari2024magnetism, siegfried2022magnetization, zhu2024geometrical, bhandari2025three, jones2024origin, riberolles2023orbital, ryan2024uncovering, gazzah2025skyrmion, gazzah2025doping, dhakal2021anisotropically, fruhling2024topological, zeng2022large, wang2024unveiling, kabir2022unusual, arachchige2022charge, pokharel2022highly, rosenberg2022uniaxial, thomas2025unusual}. Generally, the $R$166 compounds crystallize in the HfFe$_6$Ge$_6$-type structure [Fig. \ref{Structure 1}(a)] with hexagonal symmetry (space group P6/mmm), in which the $T$ atoms form a regular kagome network.

However, several members of this family, particularly the $R$Fe$_6$Ge$_6$ compounds, form superstructures derived from an average hexagonal structure with lattice parameters ($a$, $c$)$_h$. When transformed into the orthohexagonal setting, where the axes are redefined as ($c$, $\sqrt{3}a$, $a$)$_h$, these superstructures adopt distinct forms depending on the specific rare-earth element \cite{oleksyn1997crystal, schobinger1998atomic, schobinger1998fe, zaharko1999magnetic, larson2004effect}. Compounds with $R$ = Y, Tb, Dy, and Ho crystallize in the TbFe$_6$Sn$_6$-type structure with orthorhombic symmetry (space group Cmcm) as depicted in Fig. \ref{Structure 1}(b). The emergence of this superstructure originates from the ordered arrangement of $R$ and Ge atoms within the Fe–Ge host lattice. Consequently, these materials exhibit intricate structural characteristics, featuring a slightly distorted kagome network of Fe atoms within the crystallographic $a$-plane as presented in [Figs. \ref{Structure 1}(b,d)], and chains of $R$–Ge dimers oriented perpendicular to the kagome nets as noted in [Figs. \ref{Structure 1}(b,c)].

\begin{figure*}[htbp] 
    \centering
    \includegraphics[width=1.0\textwidth]{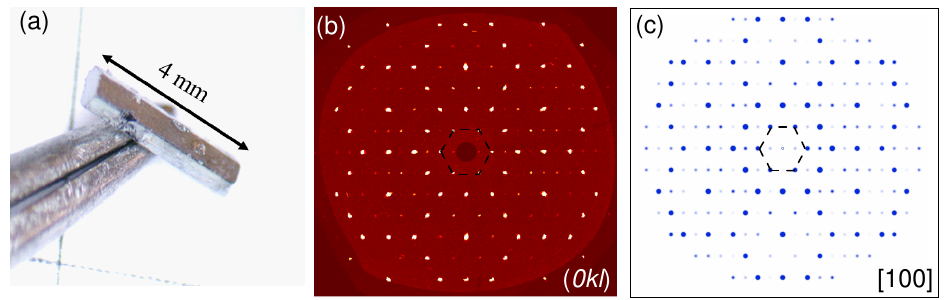}
     \caption{\small a) Optical microscope image of a representative TbFe$_6$Ge$_6$ single crystal mounted on the tip of tweezers. b) Reciprocal space map of the $(0kl)$ scattering plane obtained via single-crystal X-ray diffraction, confirming the in-plane crystallographic order and symmetry of the DyFe$_6$Ge$_6$ structure. c) Simulated single-crystal diffraction pattern of DyFe$_6$Ge$_6$ along the [100] direction, for comparison with the experimental data shown in panel (b). The dashed hexagons in panels (b) and (c) serve as visual guides.}
    \label{Structure2}
\end{figure*}

It is helpful to compare the $R$Fe$_6$Ge$_6$-Cmcm structure with the better-studied  HfFe$_6$Ge$_6$-type structure. In both structures, Fe forms a Fe$_3$ kagome network, with its hexagonal voids filled by Ge (Ge$^h$). In  HfFe$_6$Ge$_6$, the $R$ site forms a triangular layer Hf$^t$, followed by a Ge (Ge$^h$) [see Fig. \ref{Structure 1}(a)]; these stack in the sequence Hf$^t$Ge$_2^h$-Fe$_3$Ge$^t$-Ge$_2^h$-Fe$_3$Ge$^t$-Hf$^t$Ge$_2^h$, yielding the overall composition HfFe$_6$Ge$_6$.

The $R$Fe$_6$Ge$_6$-Cmcm structure is derived from another hexagonal lattice, $R$Fe$_3$Ge$_2$ as shown in Fig. \ref{Structure 1}(d). This parent structure has the same symmetry group as HfFe$_6$Ge$_6$, but consists of only two layers: $R^t$Ge$_2^h$-Fe$_3$. The $R$Fe$_6$Sn$_6$-Cmcm structure emerges when the prototype unit cell of $R$Fe$_3$Ge$_2$ is doubled along the hexagonal $c$-axis, and every other $R$ atom is replaced by a tightly bound Ge$_2$ dimer (bond length $\sim2.4$ {\AA}), yielding the overall composition of $R$Fe$_6$Ge$_6$. This produces $R$Ge$_2$ chains that penetrate the hexagonal channels of the kagome network, as illustrated in Fig. \ref{Structure 1}(f). However, because $R$ and Ge$_2$ have different shapes, it is sterically unfavorable for all $R$s to lie in one plane, and all Ge$_2$s dimers in another. Instead, they adopt a staggered arrangement within the hexagonal basal plane (which becomes $bc$ plane in Cmcm notation). It is well known that a triangular plane cannot be partitioned 1:1 without breaking the  hexagonal symmetry; accordingly, the structure becomes orthorhombic. The precise pattern of intraplanar ordering of $R$ and Ge$_2$ is more subtle. Experimentally, these sites are often disordered (although the 1D order inside each chain is strictly maintained). 

The Cmcm $R$Fe$_6$Ge$_6$ compounds have previously been investigated in polycrystalline form for their structural, magnetic and transport properties \cite{oleksyn1997crystal, schobinger1998atomic, schobinger1998fe, zaharko1999magnetic, larson2004effect}. Here, we report the single-crystal growth of three members of this family with $R$ = Y, Tb, and Dy, and present a comprehensive study of their structural, magnetic, transport, and electronic properties. Single crystal x-ray diffraction indicates that the samples are well ordered including the intraplanar $R$ and Ge$_2$ chains, which adopt a specific (and unusual) double-stripe arrangement. In all three compounds Fe orders antiferromagnetically above 400 K. Tb and Dy exhibit ferromagnetic-like ordering below 20 K, whereby Dy shows two low-temperature magnetic transitions with a complex magnetic character. All three compounds are metallic and each displays a distinct and intricate magnetoresistance. Analysis of low-temperature heat capacity of YFe$_6$Ge$_6$ reveals an unusually large Sommerfeld coefficient. Consistent with the ferromagnetic ordering in the planes, band-structure calculations for nonmagnetic YFe$_6$Ge$_6$ observe a large density of states at the Fermi level. DFT calculations also find the experimentally reported antiferromagnetic configuration to be energetically favorable; however, the resulting density of states in the magnetically ordered state is substantially reduced, more than four times smaller than expected based on the measured Sommerfeld coefficient. This behavior is reminiscent to the so-called ``d-electron heavy fermion behavior'', often observed in materials close a magnetic instability and ascribed to spin fluctuation\cite{dHF1,dHF2,dHF3}. Our material is not close to a quantum critical point, but it is worth noting that such a behavior was recently reported in another strongly ordered compound\cite{Tan}, and tentatively ascribed to soft magnetic moments and longitudinal spin fluctuations. Observation of a similar trend on YFe$_6$Ge$_6$ is interesting and deserves further studies. Unfortunately, in the Tb and Dy analogs the onset of low-temperature magnetic order prevents reliable extraction of  Sommerfeld coefficients. 

The combination of structural features, particularly the superstructure and orthorhombic distortion, with the large density of states at the Fermi level and enhanced Sommerfeld coefficient makes these compounds interesting for further study. In particular, tuning the Fermi level through chemical substitution may provide a promising route to explore the potential Fermi-surface instabilities in this family.

\begin{table*}[htbp]
\centering
\caption{Crystallographic parameters for {$R$Fe$_6$Ge$_6$} compounds ($R$ = Tb, Dy, Y) at 250(2) K using Mo K$\alpha$ radiation. Note that the kagome plane is $bc$ in this setting.}
\begin{tabular}{lccc}
\hline
\hline
Crystal system & \multicolumn{3}{c}{Orthorhombic} \\
Space group & \multicolumn{3}{c}{$Cmcm$} \\
Temperature (K) & \multicolumn{3}{c}{250(2)} \\
Wavelength (Å) & \multicolumn{3}{c}{0.71073} \\
Z (formula units) & \multicolumn{3}{c}{2} \\
2$\theta_{\min}$ (°) & \multicolumn{3}{c}{4.61} \\
2$\theta_{\max}$ (°) & \multicolumn{3}{c}{56.59} \\
Index ranges & \multicolumn{3}{c}{$-10 \leq h \leq 10$, $-23 \leq k \leq 23$, $-6 \leq l \leq 6$} \\

\text{Compounds} & 
\text{YFe\textsubscript{6}Ge\textsubscript{6}} & \text{TbFe\textsubscript{6}Ge\textsubscript{6}} & \text{DyFe\textsubscript{6}Ge\textsubscript{6}} \\
\hline
Formula weight & 1719.10 & 1859.12 & 1866.28 \\
$a$ (Å) & 8.1134(3) & 8.1379(3) & 8.1193(2) \\
$b$ (Å) & 17.6662(7) & 17.6975(6) & 17.6699(6) \\
$c$ (Å) & 5.11650(10) & 5.1257(2) & 5.11910(10) \\
Volume (Å\textsuperscript{3}) & 733.36(4) & 738.21(5) & 734.42(3) \\
Density (calc., g/cm\textsuperscript{3}) & 7.785 & 8.364 & 8.439 \\
$\mu$ (Mo K$\alpha$) (cm$^{-1}$) & 433.73 & 447.68 & 455.43 \\
Goodness of fit on $F^2$ & 1.179 & 1.132 & 1.347 \\
$R(F)$ for $F^2 > 2\sigma(F^2)$\textsuperscript{a} & 0.0155 & 0.0268 & 0.0245 \\
$R_w(F^2)$\textsuperscript{b} & 0.0447 & 0.0839 & 0.0559 \\
\hline
\hline
\end{tabular}

\vspace{1em}
\raggedright
\textsuperscript{a} $R(F) = \sum ||F_o| - |F_c|| / \sum |F_o|$ \\
\textsuperscript{b} $R_w(F^2) = \left[\sum w(F_o^2 - F_c^2)^2 / \sum w(F_o^2)^2 \right]^{1/2}$
\end{table*}

\section{Methods}\label{sec:2}

Single crystals of $R$Fe$_6$Ge$_6$ ($R$ = Y, Tb, Dy) were grown using the self-flux method. Pieces of Y, Tb, or Dy (Thermo Scientific, 99.99\%), Fe powder (Thermo Scientific, 99.999\%), and Ge pieces (Thermo Scientific, 99.999\%) were mixed in a molar ratio of 1:10:10, placed in an alumina crucible, and sealed in a fused silica ampoule under a vacuum of 10$^{-3}$ Pa. The sealed ampoule was heated in a box furnace to 1200 $^{\circ}$C over 10 hours, held at that temperature for 24 hours, then cooled to 1000 $^{\circ}$C over 35 hours. It was subsequently reheated to 1050 $^{\circ}$C over 10 hours, and without holding, cooled to 920 $^{\circ}$C over 72 hours. At 920 $^{\circ}$C, the ampoule was centrifuged to separate the crystals from the remaining molten flux. Several well-faceted, long needle-like crystals (see Fig. \ref{Structure2}(a)) weighing up to 7 mg were obtained. The typical crystal dimensions are approximately 4 mm $\times$ 0.5 mm $\times$ 0.5 mm.

Single-crystal X-ray diffraction was performed to determine the crystal structure of all three $R$Fe$_6$Ge$_6$ compounds. Data were collected on a silver-colored, block-shaped crystal with dimensions of $0.073\times0.068\times0.026$ mm$^{3}$, using a Bruker Quest diffractometer equipped with a Bruker PHOTON III detector, and employing a combination of $\omega$- and $\phi$-scans with a step size of 0.5 degree \cite{apex4}. The diffraction data were corrected for absorption and polarization effects and analyzed to determine the appropriate space group \cite{krause2015}. Structure solution was carried out using dual-space algorithms, followed by routine expansion of the initial model. Final structural refinement was performed using full-matrix least-squares analysis against $F^2$ values for all observed reflections \cite{sheldrick2015shelxt}. All non-hydrogen atoms were refined with anisotropic atomic displacement parameters, and hydrogen atoms, unless otherwise noted, were placed in calculated positions. The extinction coefficient was also refined. Following data collection, a high-precision image of the $(0kl)$ plane was captured [see Fig. \ref{Structure2}(b)].

DC magnetization, resistivity, and heat capacity measurements were carried out using two separate Quantum Design Dynacool Physical Property Measurement Systems (PPMS) equipped with 9 T and 14 T magnets. The ACMS II option was used for DC magnetization measurements. Single crystals of $R$Fe$_6$Ge$_6$ were polished to appropriate dimensions for electrical transport measurements. Crystal orientation along the [100], [010], and [001] directions was determined using a Photonic Science X-ray Laue diffractometer to align the magnetic field along the $a$-, $b$-, and $c$-axes, respectively. Resistivity and Hall effect measurements were carried out using the standard four-probe method. Platinum wires (25 $\mu$m in diameter) were used for electrical contacts, with contact resistances below 20 $\Omega$. The electrical contacts were affixed using Epotek H20E silver epoxy, and a 2 mA current was applied during transport measurements. To correct for contact misalignment, magnetoresistance and Hall resistivity data were symmetrized and anti-symmetrized, respectively, following the procedure described in Ref. \cite{bhandari2025tunable}. All magnetic and magnetotransport data presented here were measured on multiple samples and found to be reproducible (see appendix for further data).

First-principles calculations were performed using the density functional theory (DFT) based on the WIEN2k code \cite{wien} using the augmented plane-wave basis set defined by the parameter $RK_{\text{max}}=7.0$ and a $8\times8\times12$ mesh of $k$-points for Brillouin zone integration in a primitive two-formula-unit unit cell. Exchange and correlation effects were treated within the  Perdew-Burke Ernzerhof gradient-corrected local density approximation \cite{PBE}.

\section{Results and Discussion}\label{sec:3}

\subsection{Crystal Structure}\label{sec:3a}

\begin{figure*}[htbp] 
    \centering
    \includegraphics[width=1.0\textwidth]{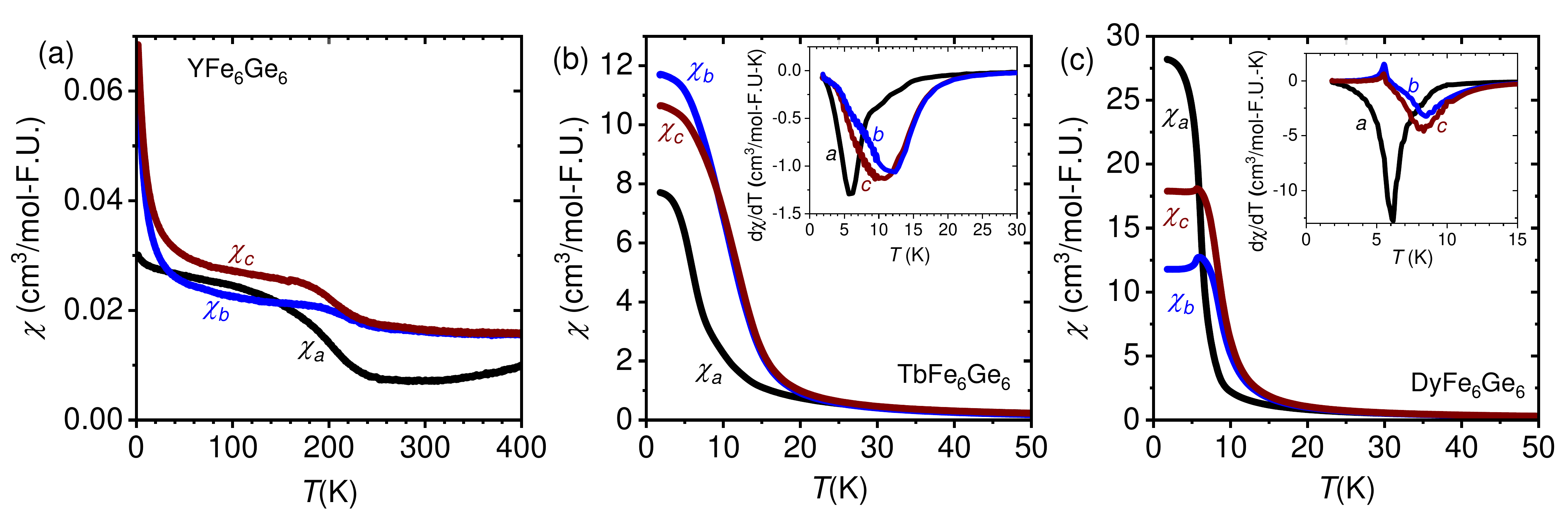}
    \caption{\small Temperature dependence magnetic susceptibility (FC) measured at 0.1 T magnetic field along $a$ (black line), $b$ (blue line), and $c$ (brown line) crystallographic axes for a) YFe$_6$Ge$_6$, b) TbFe$_6$Ge$_6$, and c) DyFe$_6$Ge$_6$. Insets in panels $b$ and $c$ show first derivative of corresponding FC magnetic susceptibility in all three crystallographic directions.}
    \label{MT}
\end{figure*}

The compounds $R$Fe$_6$Ge$_6$ crystallize in an orthorhombic structure characterized by the space group \textit{Cmcm} (No. 63). At room temperature, the lattice parameters for different $R$ atoms have been determined as follows: for $R$ = Y, a = 8.1134(3)Å, b = 17.6662(7)Å, c = 5.11650(10)Å;  $R$ = Dy, a = 8.1193(2)Å, b = 17.6699(6)Å, c = 5.1191(10)Å; and $R$ = Tb a = 8.1379(3)Å, b = 17.6975(6)Å, c = 5.1257(2)Å. A detailed summary of the single-crystal data and structural refinement parameters are provided in Table I, while fractional atomic coordinates and equivalent isotropic displacement parameters are listed comprehensively in Table II for YFe$_6$Ge$_6$, and in appendix Table III, IV for TbFe$_6$Ge$_6$ and DyFe$_6$Ge$_6$, respectively. Structural refinement parameters for disorder model are illustrated in appendix Table VI-VIII.

\begin{table}[htbp]
\caption{Atomic coordinates, occupancy, displacement parameters and Wyckoff sites for YFe$_6$Ge$_6$ compounds.}
\begin{tabular}{l@{\hskip 12pt}r@{\hskip 12pt}r@{\hskip 12pt}r@{\hskip 12pt}r@{\hskip 12pt}r@{\hskip 12pt}l}
\toprule
\text{Atom} & \text{Site} & \text{Occ.} &  $x$ & $y$ & $z$ & \text{U($_{eq}$)} \\
\midrule
Y(1)    & $4c$ & 1.000 & 0.00000 & 0.62710 & 0.75000 & 0.008 \\
Ge(1)   & $8g$ & 1.000 & 0.34708 & 0.62470 & 0.75000 & 0.008 \\
Ge(2)   & $4c$ & 1.000 & 0.00000 & 0.54202 & 0.25000 & 0.007 \\
Ge(3)   & $4c$ & 1.000 & 0.50000 & 0.70505 & 0.25000 & 0.007 \\
Ge(4)   & $4c$ & 1.000 & 0.50000 & 0.78957 & 0.75000 & 0.007 \\
Ge(5)   & $4c$ & 1.000 & 0.50000 & 0.45955 & 0.75000 & 0.007 \\
Fe(1)   & $8e$ & 1.000 & 0.24959 & 0.50000 & 0.50000 & 0.006 \\
Fe(2)   & $8g$ & 1.000 & 0.25036 & 0.62496 & 0.25000 & 0.006 \\
Fe(3)   & $8d$ & 1.000 & 0.75000 & 0.75000 & 0.50000 & 0.006 \\
\bottomrule
\end{tabular}
\end{table}

The orthorhombic unit cell of $R$Fe$_6$Ge$_6$ comprises four formula units ($Z$ = 4), totaling four $R$ atoms, twenty four (Fe) atoms, and twenty four (Ge) atoms. Specifically, there is one crystallographically unique rare-earth atom sitting on 4$c$ Wyckoff position, five distinct germanium atoms identified as Ge1 on the $8g$ site, and Ge2, Ge3, Ge4, and Ge5 which occupy distinct $4c$ sites, and three inequivalent iron atoms identified as Fe1 on the $8e$ site, Fe2 at the $8g$ site, and Fe3 at the $8d$ site.

Previous studies on polycrystalline samples of $R$Fe$_6$Ge$_6$ compounds have reported slight atomic disorder at the $R$ ($4c$) site, characterized by a positional shift of 1/2 along the $x$-axis, as well as a similar disorder involving Ge atoms occupying $8g$ sites \cite{schobinger1998atomic, cadogan1998magnetic, cadogan2000neutron}. Considering the aforementioned literature results, we paid particular attention to examining possible disorder in our single crystal samples and therefore carried out structural refinements using three different approaches: i) assuming fully ordered Wyckoff positions, ii) allowing for 10\% disorder between the rare-earth and Ge sites, and iii) allowing for 20\% disorder. The refinement converged optimally when only the ordered atomic positions were used, yielding the lowest goodness-of-fit parameters ($R_F$ and $R_w$), consistent with the discussion above and the DFT calculation, indicating that the $R$ atoms and Ge dimers are perfectly well ordered in each 1D chain. It still leaves open the possibility of some 2D disorder between the ideal chains; to resolve this, one would need to go diffuse scattering.

The refined structural parameters are summarized in Table I (see also appendix Tables V, where the results of disorder refinement are summarized). These results indicate that our single crystals are better ordered than previously studied polycrystalline samples, reflecting a high degree of positional order within the  1D Y-Ge$_2$ chains. A precision pattern of $(0kl)$ scattering plane of DyFe$_6$Ge$_6$ obtained from single crystal X-ray diffraction is presented in Fig. \ref{Structure2}(b). For comparison, a simulated precision pattern of the same plane of DyFe$_6$Ge$_6$ is presented in Fig. \ref{Structure2}(c).

\begin{figure*}[htbp] 
    \centering
    \includegraphics[width=1\textwidth]{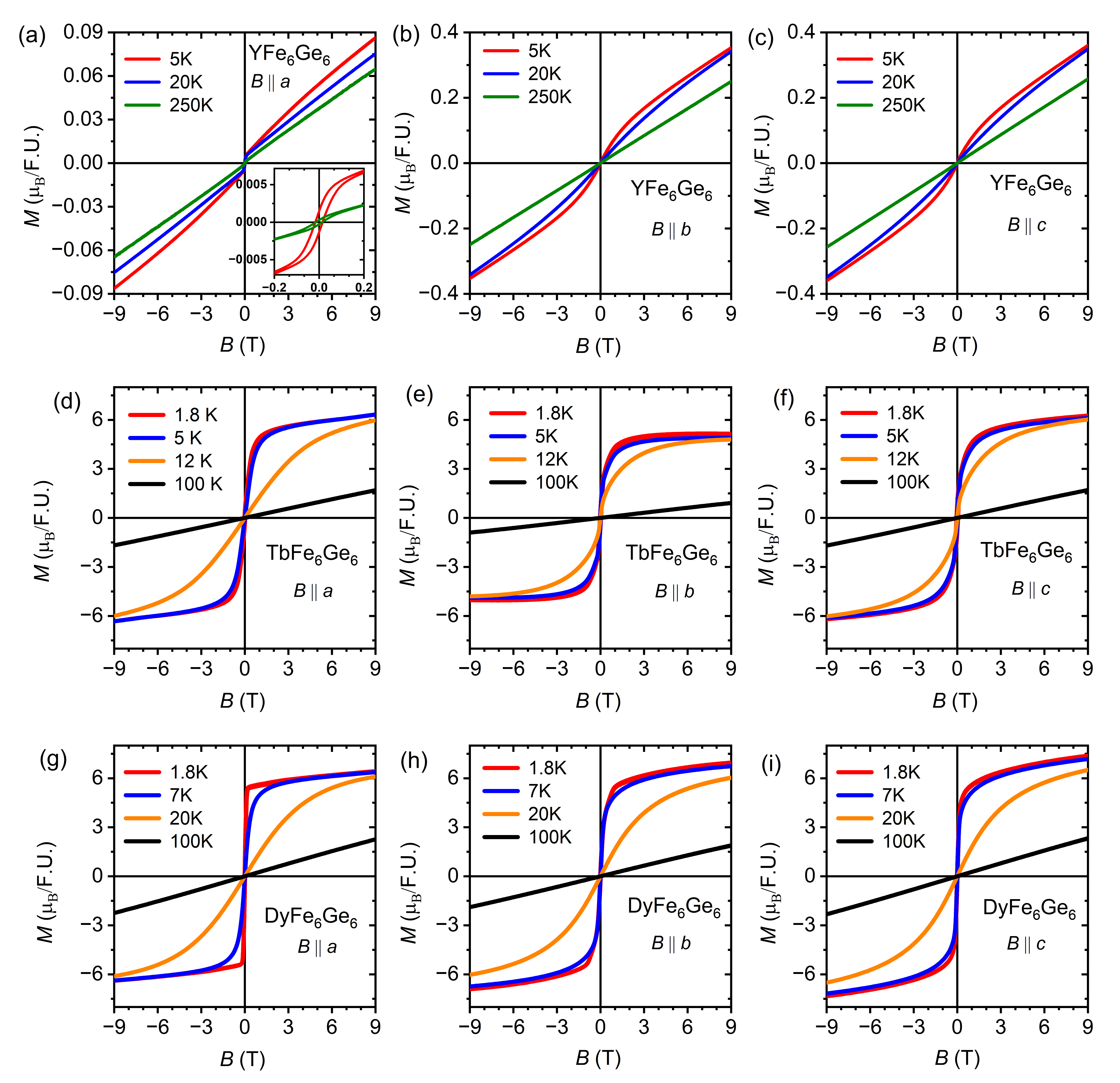}
     \caption{\small Field-dependent isothermal magnetization $M$ vs. $B$ or simply ($MB$) measured along the crystallographic $a$, $b$, and $c$ axes for (a–c) YFe$_6$Ge$_6$, (d–f) TbFe$_6$Ge$_6$, and (g–i) DyFe$_6$Ge$_6$, respectively. Measurements were performed at various temperatures with an applied magnetic field sweeping from 9 T to $-$9 T and $-$9 T to 9 T.}
    \label{MH}
\end{figure*}

\subsection{Magnetic Properties}\label{sec:3b}

Magnetism in the $R$Fe$_6$Ge$_6$ compounds has been extensively studied in polycrystalline samples. Experiments, including neutron diffraction, have established that the magnetic $R$ and Fe sublattices order independently in these materials. The Fe sublattice exhibits antiferromagnetic order, with a Néel temperature ($T_{\text{N}}$) that remains essentially constant across the series, at approximately 485 K \cite{schobinger1998fe}. Neutron diffraction measurements on powders revealed that within each Fe kagome plane, the spins are aligned ferromagnetically, while adjacent planes couple antiferromagnetically along the $a$-axis, which is also the easy axis of magnetization \cite{schobinger1998fe, schobinger1998atomic,zaharko1999magnetic}.

\begin{figure*}[htbp] 
    \centering
    \includegraphics[width=1.0\textwidth]{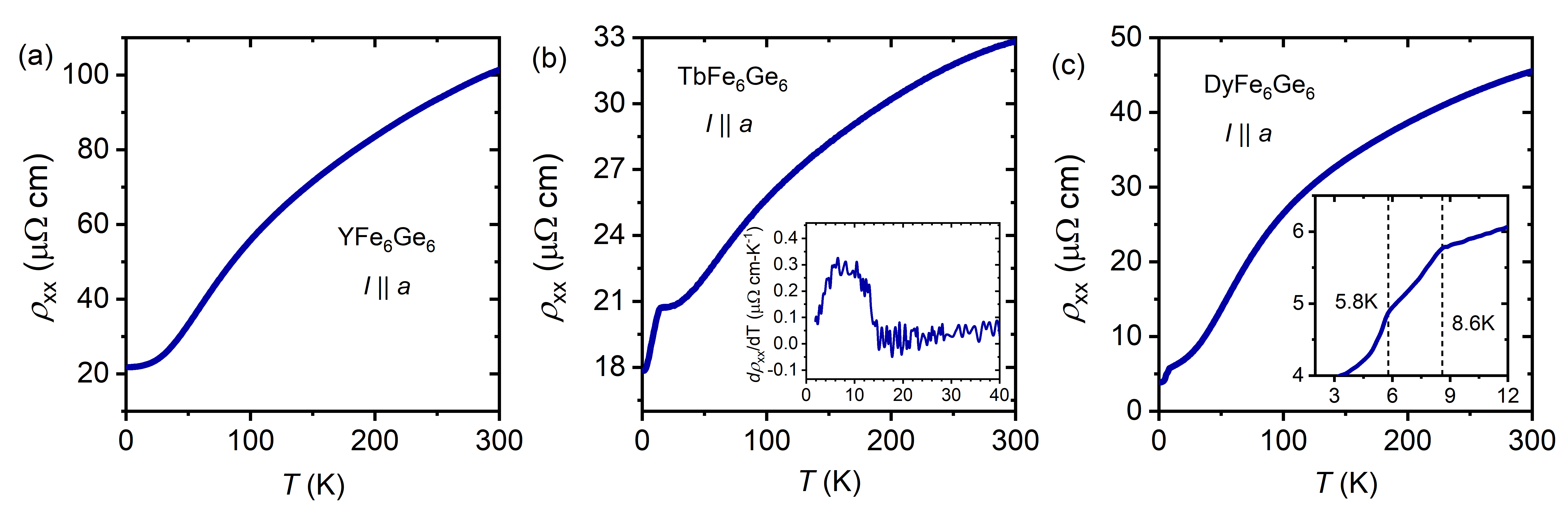}
    \caption{\small Temperature variation of longitudinal resistivity measured with current applied along $a$-axis for a) YFe$_6$Ge$_6$, b) TbFe$_6$Ge$_6$, and c) DyFe$_6$Ge$_6$. Insets in panels (b) shows the temperature derivative of the resistivity and in panels (c) shows enlarge view of low temperature data.}
    \label{RT}
\end{figure*}

The magnetic susceptibility $\chi$ of YFe$_6$Ge$_6$, measured under an external applied magnetic field $B$ of 0.1 T along all three crystallographic directions, is shown in Fig \ref{MT}(a). Since the Néel temperature ($T_{\text{N}}$) lies above the measurement range of our PPMS, no clear feature associated with antiferromagnetic (AFM) ordering is observed. However, the overall low magnitude of $\chi$ is consistent with the AFM behavior. The field-cooled (FC) and zero-field-cooled (ZFC) curves show no divergence. A slight anomaly in $\chi$ is observed between 250 K and 200 K along all directions, which is rather unusual. For $B \parallel b$ and $B \parallel c$, this feature may indicate weak spin canting. However, for $B \parallel a$, where canting is symmetry-forbidden within the measurement geometry, this behavior suggests subtle complexity in the magnetic structure. Additionally, a small upturn in $\chi$ appears below ~10 K for $B \parallel b$ and $c$, which may also be related to spin canting.

The complexity in magnetism is further observed in magnetization ( $M$ vs $B$) plots depicted in Figs. \ref{MH}(a) -\ref{MH}(c). For $B||a$, there is a sharp slope change at around 0.1 T, with a small but clear hysteresis at 5 K, which survives up to 250 K (Fig. \ref{MH}(a)). Magnetization for $B||b,c$ is rather simple. At 5 K, in each direction, a small smooth curvature is observed in $M-B$ curve, which is likely due to spin canting. The $MB$ data at higher temperatures show straight line behavior as expected in an antiferromagnet.

\begin{figure*}[htbp] 
    \centering
    \includegraphics[width=1.0\textwidth]{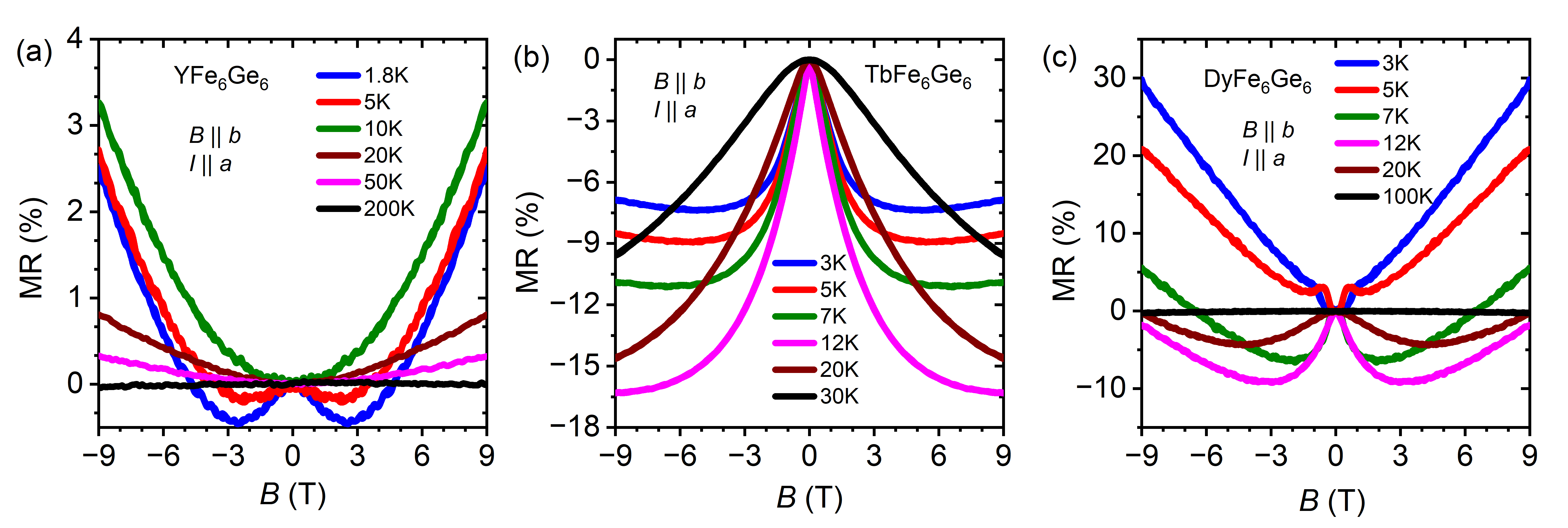}
    \caption{\small Transverse magneto-resistance (MR) of a) YFe$_6$Ge$_6$, b) DyFe$_6$Ge$_6$, and c) TbFe$_6$Ge$_6$ measured with current ($I$) applied along $a$-axis and magnetic field ($B$) along $b$-axis.}
    \label{MR}
\end{figure*}

Substituting the non-magnetic rare-earth element Y with magnetic $R$ atoms does not significantly affect the Fe magnetic ordering; neither its transition temperature nor its magnetic structure. However, the $R$ atoms themselves order magnetically at much lower temperatures. In TbFe$_6$Ge$_6$, Tb orders at 9 K, forming canted ferromagnetic layers in the $a$-plane that are stacked along the $a$-axis. This results in a net ferromagnetic component along the $a$-axis and an antiferromagnetic component along the $c$-axis \cite{schobinger1998atomic,schobinger1998fe}, indicating that the easy axis is canted away from the $a$-axis toward the $c$-axis. In DyFe$_6$Ge$_6$, Dy orders below 7.5 K, and the magnetocrystalline anisotropy shifts such that the easy-axis is now canted away from the $c$-axis toward the $b$-axis, yielding a ferromagnetic component along $c$ and an antiferromagnetic component along $b$. This variation in magnetocrystalline anisotropy between the Tb and Dy compounds is attributed to higher-order (fourth and sixth) crystal field terms of the $R^{3+}$ ions \cite{cadogan2000neutron}.

The magnetic susceptibility of TbFe$_6$Ge$_6$ is shown in Fig. \ref{MT}(b). The susceptibility increases below about 20 K in each direction. However, there is no clear peak in any of them. The temperature derivative $d\chi/dT$, however, signals a broad decrease below 20 K, with an upturn at 5, and 9 K along $a$- and $b$,$c$-axes, respectively. Magnetization measurements shown in Fig. \ref{MH}(d-f) show ferromagnetic behavior in each direction, with a small anisotropy. At 1.8 K, the moment along the $b$-axis shows saturation with a slightly smaller value (5 $\mu_{\text{B}}$/F.U.) compared to the moment along the $a$- and $c$-axes.

Likewise, magnetic susceptibility of DyFe$_6$Ge$_6$ is shown in Fig. \ref{MT}(c). $\chi_a$ increases sharply below 10 K while without any other sharp features at lower temperatures. On the other hand $d\chi_{a}/dT$, shows a sharp change at 6.2 K.  $\chi_b$ and $\chi_c$, increase rapidly below 15 K and show a marked peak near 5.5 K. Temperature derivative of these plots show an upturn at around 8.5 K and a peak at 5.5 K as shown in the inset of Fig. \ref{MT}(c). These two magnetic transitions associated with Dy ordering is in contrast to a transition observed at 7.5 K in polycrystalline samples in the previous studies \cite{cadogan2000neutron}. The magnetization data shown in Figs. \ref{MH}(g-i) show ferromagnetic behavior along all three directions at 1.8 K, with a small anisotropy along the $b$-axis. The magnetization is not linear in $B$ even at 20 K, likely due to the persistence of spin fluctuations is above the ordering temperature of Dy spins, and only demonstrate a linear behavior at 100 K, as expected for an antiferromagnet from the Fe sub-lattice. A similar temperature-dependent behavior is also seen in TbFe$_6$Ge$_6$.

\subsection{Electrical Resistivity and Magnetoresistance}\label{sec:3c}

The temperature-dependent resistivity of all three compounds, measured with current along the $a$-axis ($\rho_{xx}$ vs. $T$), is shown in Fig. \ref{RT}. All three compounds exhibit metallic behavior. The residual resistivity ratio (RRR), defined as $\rho_{xx}$(300 K)/$\rho_{xx}$(1.8 K), is 4.6 for YFe$_6$Ge$_6$, 1.8 for TbFe$_6$Ge$_6$, and 12.8 for DyFe$_6$Ge$_6$. The relatively low RRR of TbFe$_6$Ge$_6$ suggests that this compound may exhibit more disorder compared to the other two, although no such was detected in the single-crystal diffraction experiment. Interestingly, the residual resistivity of Y and Tb compounds are similar (~20 $\mu\Omega$-cm), while that of DyFe$_6$Ge$_6$ is much smaller (3.5 $\mu\Omega$-cm).

As expected, YFe$_6$Ge$_6$ shows no distinct features in $\rho_{xx}$, consistent with its only magnetic transition occurring above 400 K. In TbFe$6$Ge$_6$, a sharp drop in $\rho_{xx}$ is observed below 14 K, with a corresponding peak at 9 K in $d\rho_{xx}/dT$, consistent with the magnetic ordering of Tb atoms previously reported in polycrystalline samples \cite{schobinger1998atomic}. DyFe$_6$Ge$_6$ exhibits two transitions in resistivity, at 8.6 K and 5.8 K (see inset of Fig. \ref{RT}(c)), in agreement with the two transitions observed in the temperature derivative of magnetic susceptibility (see inset of Fig. \ref{MT}(c)).

Magnetoresistance (MR) defined by: \[
\text{MR} = \left( \frac{\rho(B) - \rho(0)}{\rho(0)} \right) \times 100\%,
\]
where $\rho(B)$ is the resistivity in an applied magnetic field $B$, and $\rho(0)$ is the resistivity measured in zero magnetic field, was measured with current along the $a$-axis and the magnetic field applied along the $b$-axis for all three compounds, as shown in Fig. \ref{MR}. At low temperatures (e.g., 1.8 and 5 K), YFe$_6$Ge$_6$ exhibits negative magnetoresistance below approximately 3 T, which then becomes positive at higher fields. At 10 K, the MR is entirely positive. A similar trend, though with reduced magnitude, is observed up to 50 K. At 200 K, a very small negative MR reappears. This magnetoresistance behavior suggests that below 10 K, there is a ferromagnetic (FM) component, likely due to spin canting, as hypothesized from the susceptibility data in Fig. \ref{MT}(a), which is suppressed by the external magnetic field, leading to negative MR. Once the FM component is fully suppressed, it becomes positive. The small negative MR observed at 200 K may have a similar origin, possibly related to the susceptibility anomaly seen around 250 K.   

The MR of TbFe$_6$Ge$_6$ (Fig. \ref{MR}(b)) is negative even up to 30 K, suggesting that fluctuation of Tb moments persists well above its ordering temperature. DyFe$_6$Ge$_6$, however, shows a very different behavior than both of Y and Tb compounds. The MR below 5 K is positive. At and above 7 K, it shows behavior similar to that of YFe$_6$Ge$_6$. This is quite strange for a system where the Dy moments are canted toward $b$-axis from $c$-axis. At low temperatures, there should be suppression of the ferromagnetic fluctuation due to the magnetic field applied along $b$-axis and one would expect negative magnetoresistance as in the case of Tb. This suggests that the magnetic structure is more complex than determined from the powder neutron diffraction, which is likely as the powder neutron diffraction does not pick up the two magnetic transitions observed in magnetic, transport and thermal (see below) measurements in single crystals of TbFe$_6$Ge$_6$.

\begin{figure*}[htbp] 
    \centering
    \includegraphics[width=0.9\textwidth]{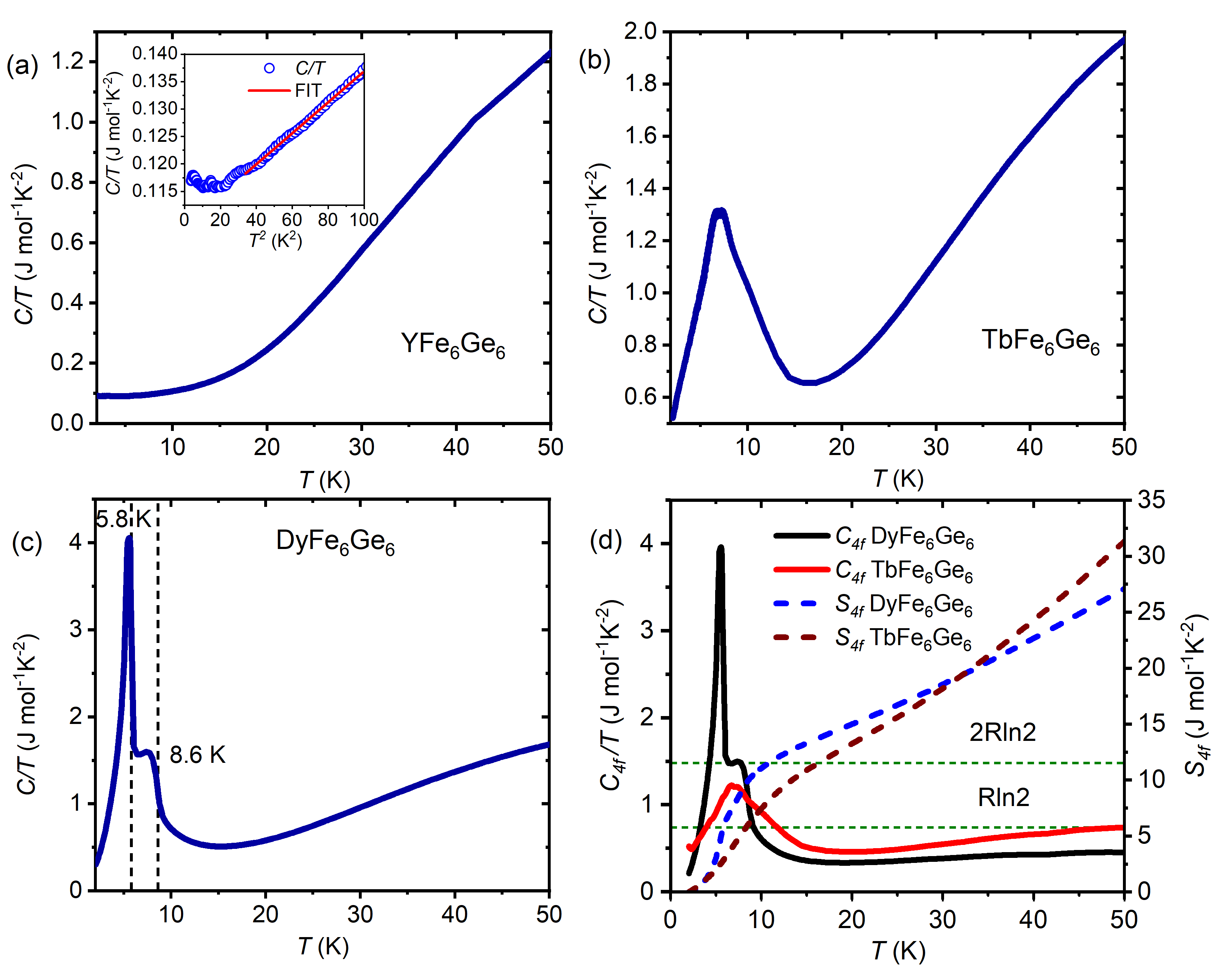}
     \caption{\small Specific heat capacity divided by the temperature $C/T$ of a) YFe$_6$Ge$_6$ (inset shows linear fit of $C/T$ vs. $T^2$), b) TbFe$_6$Ge$_6$, and c) DyFe$_6$Ge$_6$.  d) 4f contribution to the heat capacity $C$$_{4\text{f}}$($T$) of DyFe$_6$Ge$_6$ and TbFe$_6$Ge$_6$ (Left panel, straight line), and temperature dependence of the 4f contribution to the entropy $S$$_{4\text{f}}$($T$) (right axis, dashed lines). Dotted lines represent extrapolation to temperature $T=0$.}
    \label{HC}
\end{figure*}

\subsection{Heat Capacity}\label{sec:3d}

\begin{figure*}[htbp] 
    \centering
    \includegraphics[width=1.0\textwidth]{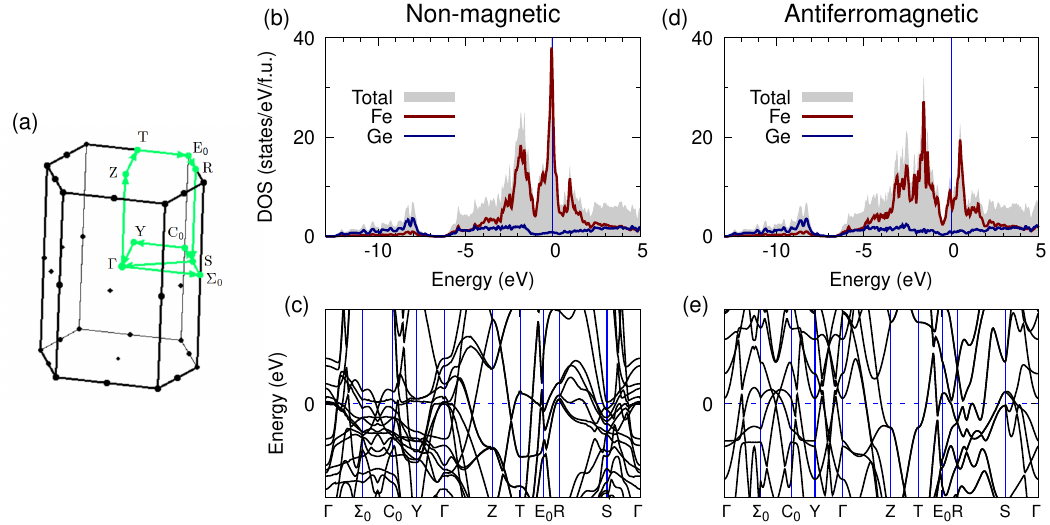}
    \caption {\small \textbf{First-principles electronic structure of YFe$_6$Ge$_6$}. (a) The first Brillouin zone with highlighted high-symmetry points. (b, c) The density
    of states (DOS) and the band structure of YFe$_6$Ge$_6$ from a non-magnetic DFT calculation.
    (d,c) The DOS and the band structure of YFe$_6$Ge$_6$ from an antiferromagnetic DFT calculation. This antiferromagnetic phase corresponds to ferromagnetic kagome Fe planes that are coupled antiferromagnetically.}
    \label{fig_dft}
\end{figure*}

The heat capacity of the RFe$_6$Ge$_6$ compounds (R = Y, Dy, Tb), plotted as $C/T$ vs. $T$, is shown in Fig. \ref{HC}. As depicted in Fig. \ref{HC}(a), YFe$_6$Ge$_6$ does not exhibit any anomaly, consistent with the high-temperature antiferromagnetic ordering of the Fe moments. The electronic contribution to the heat capacity, $\gamma$, obtained from a linear fit to $C/T$ vs. $T^2$ between 6 and 10 K using the expression $C/T = \gamma + \beta T^2$ [inset of Fig. \ref{HC}(a)], produces a value of 84 mJ·mol-F.U.$^{-1}$·K$^{-2}$. We note an unusual flattening of the $C/T$ vs $T^2$ data below 6 K; therefore, the fit was performed only above this temperature. We measured three samples from three different growth batches (the other two datasets are presented in the appendix in Fig. 15), all giving consistent results. The density of states at the Fermi level, $N(E_{\text{F}})$, estimated from this $\gamma$ using the relation $N(E_{\text{F}}) = 3\gamma / (\pi^2 k^2_{\text{B}} N_{\text{A}})$, where $k_{\text{B}}$ is the Boltzmann constant and $N_{\text{A}}$ is Avogadro’s number, is 35.7 states·eV$^{-1}$·F.U.$^{-1}$. This is more than four times the value obtained from DFT calculations in the antiferromagnetic state (see Fig. \ref{fig_dft}(b)), but im surprisingly good agreement with the result for the non-magnetic state (see Fig. \ref{fig_dft}(a)). The phonon heat capacity coefficient, $\beta$, is 0.22 mJ·mol-F.U.$^{-1}$·K$^{-4}$. Note that had we taken the actual value of $\gamma$ at $T\rightarrow 0$ instead of extrapolating the $T\agt 6$ K data, we would get an even stronger, nearly sixfold, mass renormalization, placing this material even more firmly in the ``“d-electron heavy fermion'' category \cite{dHF1,dHF2,dHF3,Tan}. The Debye temperature, estimated from $\beta$ using the relation $\Theta_{D} = \left( \frac{12\pi^4 r R}{5\beta} \right)^{1/3}$, where $r$ is the number of atoms per formula unit and $R$ is the gas constant, is 486 K.

The $C/T$ data for TbFe$_6$Ge$_6$, shown in Fig. \ref{HC}(b), exhibits a broad lambda-like anomaly centered at 9 K, consistent with the reported magnetic ordering temperature of this compound and our susceptibility and resistivity measurements. In contrast, DyFe$_6$Ge$_6$ shows two distinct peaks in $C/T$, corresponding to the two magnetic transitions observed in susceptibility and resistivity measurements. The first peak at 8.6 K is broad and small, while the second one, at 5.8 K, is sharp and prominent. This difference in the magnetic ordering between the Tb and Dy compounds can be understood by analyzing the entropy released at the respective magnetic transition temperatures.

To isolate the magnetic contribution from the 4\textit{f} electrons, the 4\textit{f} component of the heat capacity, $C_{4f}/T$, was extracted by subtracting the corresponding $C/T$ data of the reference compound YFe$_6$Ge$_6$ with non-magnetic $R$ atom from those of TbFe$_6$Ge$_6$ and DyFe$_6$Ge$_6$ \cite{ghimire2016physical}, as shown in Fig. \ref{HC}(d), and the 4$f$-electron magnetic entropy, $S_{4f}$, was calculated by using $S_{4f}$ = $\int_0^T (C_{4f}/T')dT' $. In DyFe$_6$Ge$_6$, an entropy of $R\ln2$, corresponding to the first crystal field doublet among its eight Kramers doublets, is released just above the lower-temperature transition at 5.8 K. Only a part of the additional $R\ln2$ is released just above the 8.6 K transition. This indicates that one low-lying doublet is involved in the magnetic ordering below 5.5 K while the other double is partially involved in the transitions at 8.6 K.

While Dy$^{3+}$ is a Kramers ion and thus always has doublet states, Tb$^{3+}$ is a non-Kramers ion and does not possess symmetry-protected doublets. However, the observation that $R\ln2$ entropy is released just above the Tb magnetic ordering temperature suggests the presence of a quasi-doublet. This quasi-doublet likely leads to the reduced and broadened heat capacity anomaly observed in TbFe$_6$Ge$_6$, similar to the partially filled second double in DyFe$_6$Ge$_6$.

\subsection{First-principles calculations}\label{sec:3d}

Now, let us focus on the properties of the Fe magnetic sublattice of $R$Fe$_6$Ge$_6$. Therefore,  we will use the results obtained for YFe$_6$Ge$_6$, where Y is non-magnetic, to present and discuss our findings. We found that the non-magnetic (NM) DFT state of YFe$_6$Ge$_6$ is higher in energy by about 250~meV per Fe compared to a magnetically-polarized state with ferromagnetically (FM)
ordered Fe moments. Between the two considered magnetically-polarized states, the FM one and an antiferromagnetic (AFM) one with ferromagnetic kagome Fe planes coupled antiferromagnetically,
the latter has a lower energy by 20~meV per Fe. This is in agreement with the experimentally determined magnetic ground state of the Fe sublattice in $R$Fe$_6$Ge$_6$. We find the Fe magnetic moment to be very close to 2~$\mu_{\text{B}}$. The considerable energy lowering upon spin polarization can be related to the sharp peak in the NM density of states (DOS) [Fig. \ref{fig_dft}(b)] at the Fermi level, $E_{\text F}$, which is substantially reduced in the spin-polarized phases [Fig. \ref{fig_dft}(d)]. It is important to note that, as one can see in Figs. \ref{fig_dft}(b,d), the states near the Fermi level are predominantly of Fe character. Near the Fermi-energy $E_{\text F}$ the band structure of $R$Fe$_6$Ge$_6$ as well as their Fermi surface remain complicated and certainly three-dimensional even in the magnetic ground state, as suggested by comparing the NM and AFM band structures in Figs. \ref{fig_dft}(c,e).

Total energy calculations elucidate the origin of the intriguing orthorhombic distortion described above. DFT calculations find several different energy scales. First, introducing disorder along a one-dimensional (1D) $a$-axis chain (...Y-Ge$_2$-Y-Ge$_2$...) incurs a very large energy cost of the order of 1~eV, indicating that each individual chain can be assumed to be perfectly ordered. Second, the relative ordering of these chains within the $bc$ plane is governed by a much smaller energy scale (of the order of 1~meV). Consequently, the problem can be effectively mapped onto a two-dimensional triangular Ising model (TIM), in which the spin states correspond to the presence of either Y or a Ge$_2$ dimer at a given site within a particular $bc$ plane.

\begin{figure}[h!] 
    \centering
    \includegraphics[width=.45\textwidth]{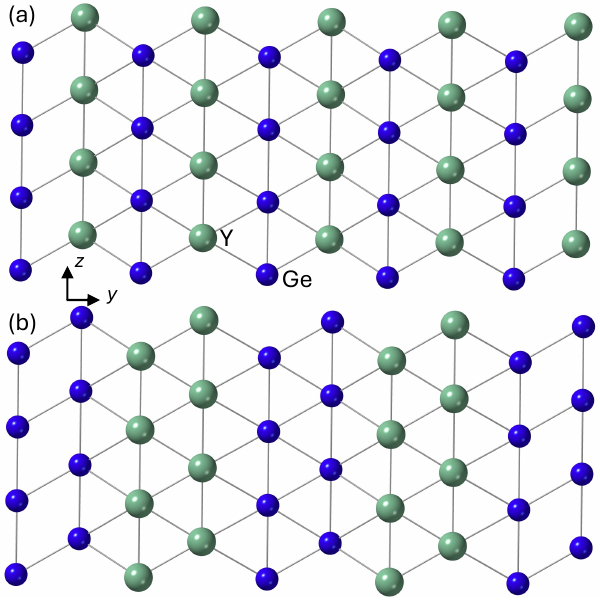}

    \caption {\small \textbf{Triangular Ising Model.} (a) 1D ferromagnetic chain structure of Y and Ge$_2$ dimers. (b) Double stripe structure of Y and Ge$_2$ dimers.} 
    \label{fig_Is}
\end{figure}

An antiferromagnetic interaction in this Ising system corresponds to an elastic energy penalty reflecting the preference of Y atoms to neighbor Ge$_2$ dimers, whereas a ferromagnetic interaction corresponds to a preference of the like neighbors. The ground state of the nearest-neighbor TIM (with the zero net magnetization) consists of one-dimensional ferromagnetic chains ordered along one of the three hexagonal directions, as illustrated in Fig. \ref{fig_Is}(a). However, the experimentally observed ``double stripe'' arrangement shown in Figs. \ref{Structure 1}(b), and \ref{fig_Is}(b), is not favored within this nearest-neighbor approximation. Given that elastic interactions are intrinsically long-ranged, it is not surprising that the double stripe configuration can emerge as the ground state (specifically, it is realized when the nearest neighbor interaction is ``ferroelastic'' and the second-neighbor interaction is ``antiferroelastic'',
estimated from our calculations to be close to -1.8~meV and 8~meV, respectively). Thus, the DFT calculations suggest the possibility of a considerable intraplanar disorder coexisting with perfect ordering along the $a$-axis and a well defined orthorhombic alignment. Experimentally resolving such a state would require diffuse X-ray scattering measurements and can be part of future investigations.

In future studies, it would be interesting to use first-principles calculations to explore in detail the rare-earth magnetism in $R$Fe$_6$Ge$_6$, taking advantage, for example, of the latest developments in DFT methodologies for treating strong correlation effects in $f$-electron ions\cite{Lee_2025}. Also, even though the experimentally determined crystal structures of $R$Fe$_6$Ge$_6$ are found to be stable and close to the theoretical equilibrium, positional disorder involving $R$ atoms can potentially strongly affect magnetic properties at low temperatures and therefore deserves further investigation.

\section*{V. Conclusion}\label{sec:V}

In conclusion, single crystals of $R$Fe$_6$Ge$_6$ (where R= Y, Dy, and Tb) are successfully synthesized, all crystallizing in an orthorhombic structure characterized by distorted kagome layers. The magnetic behavior of these compounds has been systematically investigated through magnetic susceptibility, isothermal magnetization, and magnetotransport measurements along different crystallographic directions.
YFe$_6$Ge$_6$, which lacks of 4f electrons, exhibits a canted antiferromagnetic ordering. The magnetoresistance (MR) response in this compound reveals a negative MR both at low and high temperatures, indicating suppression of spin fluctuations in the system.

In contrast, TbFe$_6$Ge$_6$ demonstrates ferromagnetic behavior at low temperatures, as evidenced by both magnetic and transport measurements. DyFe$_6$Ge$_6$, on the other hand, displays a more complex magnetic behavior, likely due to the presence of partially filled 4f orbitals. The system exhibits two magnetic transitions at 5.8 K and 8.6 K, reflecting the coexistence of ferromagnetic and antiferromagnetic interactions. The MR behavior further corroborates this complexity, showing a positive MR below the Dy ordering temperature and a transition to negative MR above it, likely due to competing spin configurations. The progression from canted antiferromagnetism in YFe$_6$Ge$_6$ to ferromagnetism in TbFe$_6$Ge$_6$ and finally to a set of complex magnetic phases in DyFe$_6$Ge$_6$, highlights the critical role of the 4f electron configuration in tuning the magnetic ground state of the RFe$_6$Ge$_6$ family.

\section*{ ACKNOWLEDGMENTS}
N.J.G acknowledges the support from the NSF CAREER award  DMR-2343536. I. I. M. was supported by the National Science Foundation award No. DMR-2403804. M.P.G. acknowledges the University Grants Commission, Nepal for the financial support through Collaborative Research Grants (award no. CRG-78/79 S\&T-03). 
 

\section*{References}

%

\appendix
\onecolumngrid

\section*{Appendix: Additional Data}

The reproduced datasets of magnetization, electrical resistivity, magnetoresistance, and heat capacity for the $R$Fe$_6$Ge$_6$ (R = Y, Tb, Dy) series, obtained from a newly synthesized batch of single crystals grown independently, exhibit excellent agreement with previously reported experimental results. This consistency confirms the high reproducibility and reliability of the growth process as well as the stability of the intrinsic magnetic and electronic properties across different sample batches.
To reproduce and reverify the unusual behavior of specific heat of YFe$_6$Ge$_6$ in $C/T$ vs. $T$$^2$, multiple batch of sample has been grown and in Fig. \ref{Y166 HC Reproduced}(a,b) confirms its reproducibility in different batch of samples and from batch 2 sample 2 $\gamma$ value was found to be 94 mJ·mol-F.U.$^{-1}$·K$^{-2}$ and from batch 3 sample 3 it was found to be 99 mJ·mol-F.U.$^{-1}$·K$^{-2}$. Similarly $\beta$ was calculated as 0.27 mJ·mol-F.U.$^{-1}$·K$^{-4}$ and 0.25 mJ·mol-F.U.$^{-1}$·K$^{-4}$ respectively which matches well with the previously calculated values.

\begin{figure*}[htbp] 
    \centering
    \includegraphics[width=1\textwidth]{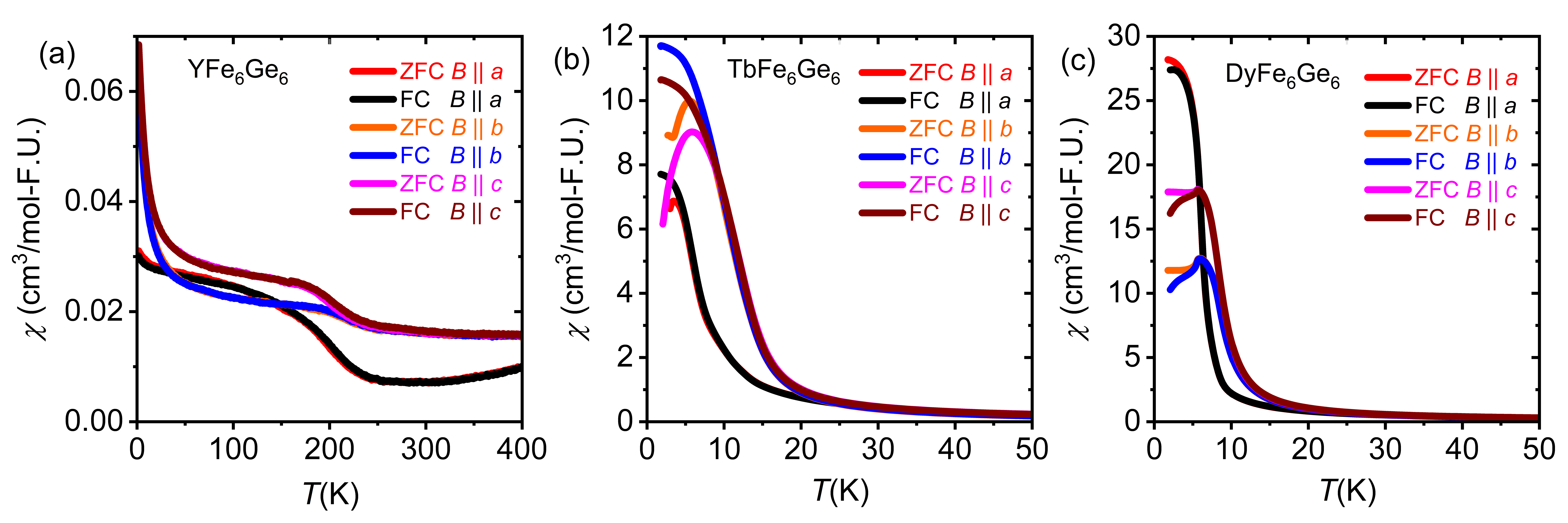}
    \caption{\small Temperature dependence magnetic susceptibility (FC, ZFC) measured at 0.1 T magnetic field in all three crystallographic directions for a) YFe$_6$Ge$_6$, b) TbFe$_6$Ge$_6$ and c) DyFe$_6$Ge$_6$.}
    \label{MT ALL}
\end{figure*}

\begin{figure*}[htbp] 
    \centering
    \includegraphics[width=1.0\textwidth]{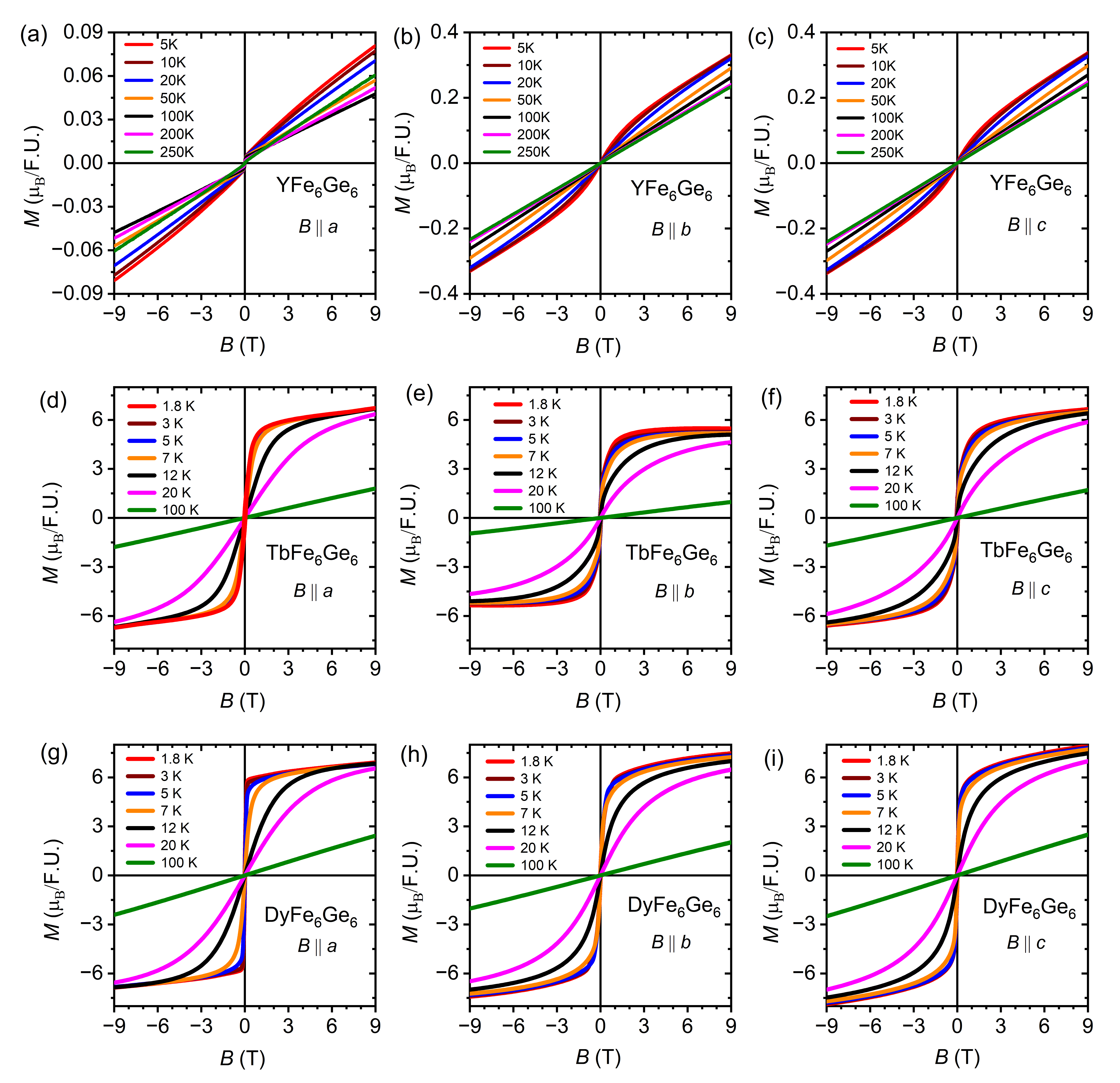}
    \caption{\small Field-dependent isothermal magnetization ($M$ vs. $B$) measurements were carried out along the crystallographic $a$, $b$, and $c$ axes for three compounds: (a–c) YFe$_6$Ge$_6$, (d–f) TbFe$_6$Ge$_6$, and (g–i) DyFe$_6$Ge$_6$. Data were reproduced using a different batch of single-crystalline samples to ensure consistency and reproducibility. The measurements were performed at selected temperatures under an external magnetic field varying from 9 T to $-$9 T and $-$9 T to 9 T.}
    \label{MH REPRODUCED}
\end{figure*}

\begin{table}[htbp]
\caption{Atomic coordinates, occupancy, isotropic displacement parameters, and Wyckoff sites for TbFe$_6$Ge$_6$ compounds.}
\begin{tabular}{l@{\hskip 12pt}r@{\hskip 12pt}r@{\hskip 12pt}r@{\hskip 12pt}r@{\hskip 12pt}r@{\hskip 12pt}l}
\toprule
\text{Atom} & \text{Site} & \text{Occ.} & $x$ & $y$ & $z$ & \text{U($_{eq}$)} \\
\midrule
Tb(1)  & $4c$ & 1.000 & 1.00000 & 0.37278 & 0.25000  & 0.006 \\
Ge(1)  & $8g$ & 1.000 & 0.65302 & 0.37538 & 0.25000  & 0.005 \\
Ge(2)  & $4c$ & 1.000 & 1.00000 & 0.45803 & 0.75000  & 0.006 \\
Ge(3)  & $4c$ & 1.000 & 0.50000 & 0.29512 & 0.75000  & 0.006 \\
Ge(4)  & $4c$ & 1.000 & 0.50000 & 0.21055 & 0.25000  & 0.006 \\
Ge(5)  & $4c$ & 1.000 & 0.50000 & 0.54044 & 0.25000  & 0.005 \\
Fe(1)  & $8g$ & 1.000 & 0.74976 & 0.37500 & 0.75000  & 0.005 \\
Fe(2)  & $8e$ & 1.000 & 0.75066 & 0.50000 & 0.50000  & 0.005 \\
Fe(3)  & $8d$ & 1.000 & 0.25000 & 0.25000 & 0.50000  & 0.005 \\
\bottomrule
\end{tabular}
\end{table}

\begin{table}[htbp]
\caption{Atomic coordinates, occupancy, isotropic displacement parameters, and Wyckoff sites for DyFe$_6$Ge$_6$ compounds.}
\begin{tabular}{l@{\hskip 12pt}r@{\hskip 12pt}r@{\hskip 12pt}r@{\hskip 12pt}r@{\hskip 12pt}r@{\hskip 12pt}l}
\toprule
\text{Atom} & \text{Site} & \text{Occ.} & $x$ & $y$ & $z$ & \text{U($_{eq}$)} \\
\midrule
Dy(1)   & $4c$ & 1.000 & 0.00000 & 0.62729 & 0.75000 & 0.006 \\
Ge(1)   & $8g$ & 1.000 & 0.34713 & 0.62468 & 0.75000 & 0.007 \\
Ge(2)   & $4c$ & 1.000 & 0.00000 & 0.54201 & 0.25000 & 0.006 \\
Ge(3)   & $4c$ & 1.000 & 0.50000 & 0.70488 & 0.25000 & 0.007 \\
Ge(4)   & $4c$ & 1.000 & 0.50000 & 0.78947 & 0.75000 & 0.007 \\
Ge(5)   & $4c$ & 1.000 & 0.50000 & 0.45961 & 0.75000 & 0.007 \\
Fe(1)   & $8e$ & 1.000 & 0.24961 & 0.50000 & 0.50000 & 0.007 \\
Fe(2)   & $8g$ & 1.000 & 0.25046 & 0.62494 & 0.25000 & 0.007 \\
Fe(3)   & $8d$ & 1.000 & 0.75000 & 0.75000 & 0.50000 & 0.007 \\
\bottomrule
\end{tabular}
\end{table}

\begin{table*}[htbp]
\centering
\caption{Crystallographic parameters for {$R$Fe$_6$Ge$_6$} compounds ($R$ = Tb, Dy, Y) at 250(2) K using Mo K$\alpha$ radiation.}
\begin{tabular}{lcccccc}
\hline
\hline
Crystal system & \multicolumn{6}{c}{Orthorhombic} \\
Space group & \multicolumn{6}{c}{Cmcm} \\
Temperature (K) & \multicolumn{6}{c}{250(2)} \\
Wavelength (Å) & \multicolumn{6}{c}{0.71073} \\
Z (formula units) & \multicolumn{6}{c}{2} \\
2$\theta_{\min}$ (°) & \multicolumn{6}{c}{4.61} \\
2$\theta_{\max}$ (°) & \multicolumn{6}{c}{56.59} \\
Index ranges & \multicolumn{6}{c}{$-10 \leq h \leq 10$, $-23 \leq k \leq 23$, $-6 \leq l \leq 6$} \\

\text{Compounds} & 
\text{YFe\textsubscript{6}Ge\textsubscript{6} 10\%} & \text{YFe\textsubscript{6}Ge\textsubscript{6}20\%} & \text{TbFe\textsubscript{6}Ge\textsubscript{6}10\%} & 
\text{TbFe\textsubscript{6}Ge\textsubscript{6}20\%} &
\text{DyFe\textsubscript{6}Ge\textsubscript{6}10\%} &
\text{DyFe\textsubscript{6}Ge\textsubscript{6}20\%} \\
\hline
Formula weight & 1719.10 & 1719.10 & 1859.12 & 1859.12 & 1866.28 & 1866.28 \\
$a$ (Å) & 8.1134(3) & 8.1134(3) & 8.1379(3) & 8.1379(3) & 8.1193(2) & 8.1193(2) \\
$b$ (Å) & 17.6662(7) & 17.6662(7) & 17.6975(6) & 17.6975(6) & 17.6699(6) & 17.6699(6) \\
$c$ (Å) & 5.11650(10) & 5.11650(10) & 5.1257(2) & 5.1257(2) & 5.11910(10) & 5.11910(10) \\
Volume (Å\textsuperscript{3}) & 733.36(4) & 733.36(4) & 738.21(5) & 738.21(5) & 734.42(3) & 734.42(3) \\
Density (calc., g/cm\textsuperscript{3}) & 7.785 & 7.785 & 	8.364 & 8.365 & 8.439 & 8.439 \\
$\mu$ (Mo K$\alpha$) (cm$^{-1}$) & 433.73 & 433.73 & 447.68 & 447.68 & 455.43 & 455.43 \\
Goodness of fit on $F^2$ & 2.406 & 2.393 & 1.206 & 1.138 & 5.081  & 4.627 \\
$R(F)$ for $F^2 > 2\sigma(F^2)$\textsuperscript{a} & 0.1418 & 0.1659 & 0.0378 & 0.0741 & 0.2904 & 0.3162 \\
$R_w(F^2)$\textsuperscript{b} & 0.4751 & 0.4737 & 0.1132 & 0.2185 & 0.7058 & 0.6835 \\
\hline
\hline
\end{tabular}
\end{table*}

\begin{table*}[htbp]
\caption{Atomic coordinates, occupancy, isotropic displacement parameters, and Wyckoff sites for YFe$_6$Ge$_6$ compounds with 10\% and 20\% disorder.}
\begin{tabular}{l@{\hskip 12pt}r@{\hskip 12pt}r@{\hskip 12pt}r@{\hskip 12pt}r@{\hskip 12pt}r@{\hskip 12pt}r@{\hskip 12pt}r@{\hskip 12pt}r@{\hskip 12pt}l}
\toprule
\text{Atom} & \text{Site} & \text{Occ.10\%} & \text{Occ.20\%} & $x$ & $y$ & $z$ & \text{U($_{eq}$) 10\%} & \text{U($_{eq}$) 20\%} \\
\midrule
Y(1)   & $4c$ & 0.890 & 0.800 & 0.00000 & 0.62710 & 0.75000 & 0.030 & 0.037 \\
Y(1)   & $4c$ & 0.110 & 0.200 & 0.50000 & 0.62710 & 0.75000 & -0.001 & 0.016 \\
Ge(1)   & $8g$ & 0.890 & 0.800 & 0.34708 & 0.62470 & 0.75000 & 0.021 & 0.024 \\
Ge(1)   & $8g$ & 0.110 & 0.200 & 0.84708 & 0.62470 & 0.75000 & -0.037 & -0.001 \\
Ge(2)   & $4c$ & 1.000 & 1.000 & 0.00000 & 0.54202 & 0.25000 & 0.024 & 0.027 \\
Ge(3)   & $4c$ & 1.000 & 1.000 & 0.50000 & 0.70505 & 0.25000 & 0.018 & 0.022 \\
Ge(4)   & $4c$ & 1.000 & 1.000 & 0.50000 & 0.78957 & 0.75000 & 0.021 & 0.025 \\
Ge(5)   & $4c$ & 1.000 & 1.000 & 0.50000 & 0.45955 & 0.75000 & 0.015 & 0.019 \\
Fe(1)   & $8e$ & 1.000 & 1.000 & 0.24959 & 0.50000 & 0.50000 & 0.021 & 0.024 \\
Fe(2)   & $8g$ & 1.000 & 1.000 & 0.25036 & 0.62496 & 0.25000 & 0.021 & 0.025 \\
Fe(3)   & $8d$ & 1.000 & 1.000 & 0.75000 & 0.75000 & 0.50000 & 0.021 & 0.024 \\
\bottomrule
\end{tabular}
\end{table*}

\begin{table*}[htbp]
\caption{Atomic coordinates, occupancy, isotropic displacement parameters, and Wyckoff sites for TbFe$_6$Ge$_6$ compounds with 10\% and 20\% disorder.}
\begin{tabular}{l@{\hskip 12pt}r@{\hskip 12pt}r@{\hskip 12pt}r@{\hskip 12pt}r@{\hskip 12pt}r@{\hskip 12pt}r@{\hskip 12pt}r@{\hskip 12pt}r@{\hskip 12pt}l}
\toprule
\text{Atom} & \text{Site} & \text{Occ.10\%} & \text{Occ.20\%} & $x$ & $y$ & $z$ & \text{U($_{eq}$) 10\%} & \text{U($_{eq}$) 20\%} \\
\midrule
Tb(1)   & $4c$ & 0.890 & 0.800 & 1.00000 & 0.37278 & 0.25000 & 0.007 & 0.007 \\
Tb(1)   & $4c$ & 0.110 & 0.200 & 0.50000 & 0.37278 & 0.25000 & 0.044 & 0.055 \\
Ge(1)   & $8g$ & 0.890 & 0.800 & 0.65304 & 0.37538 & 0.25000 & 0.006 & 0.006 \\
Ge(1)   & $8g$ & 0.110 & 0.200 & 0.15304 & 0.37538 & 0.25000 & 0.047 & 0.046 \\
Ge(2)   & $4c$ & 1.000 & 1.000 & 1.00000 & 0.45803 & 0.75000 & 0.009 & 0.016 \\
Ge(3)   & $4c$ & 1.000 & 1.000 & 0.50000 & 0.54045 & 0.25000 & 0.009 & 0.016 \\
Ge(4)   & $4c$ & 1.000 & 1.000 & 0.50000 & 0.29510 & 0.75000 & 0.014 & 0.014 \\
Ge(5)   & $4c$ & 1.000 & 1.000 & 0.50000 & 0.21054 & 0.25000 & 0.014 & 0.015 \\
Fe(1)   & $8d$ & 1.000 & 1.000 & 0.25000 & 0.25000 & 0.50000 & 0.011 & 0.015 \\
Fe(2)   & $8g$ & 1.000 & 1.000 & 0.74975 & 0.37500 & 0.75000 & 0.010 & 0.015 \\
Fe(3)   & $8e$ & 1.000 & 1.000 & 0.75065 & 0.50000 & 0.50000 & 0.010 & 0.015 \\
\bottomrule
\end{tabular}
\end{table*}

\begin{table*}[htbp]
\caption{Atomic coordinates, occupancy, isotropic displacement parameters, and Wyckoff sites for DyFe$_6$Ge$_6$ compounds with 10\% and 20\% disorder.}
\begin{tabular}{l@{\hskip 12pt}r@{\hskip 12pt}r@{\hskip 12pt}r@{\hskip 12pt}r@{\hskip 12pt}r@{\hskip 12pt}r@{\hskip 12pt}r@{\hskip 12pt}r@{\hskip 12pt}l}
\toprule
\text{Atom} & \text{Site} & \text{Occ.10\%} & \text{Occ.20\%} & $x$ & $y$ & $z$ & \text{U($_{eq}$) 10\%} & \text{U($_{eq}$) 20\%} \\
\midrule
Dy(1)   & $4c$ & 0.890 & 0.800 & 0.00000 & 0.62729 & 0.75000 & 0.303 & 0.076 \\
Dy(1)   & $4c$ & 0.110 & 0.200 & 0.50000 & 0.62729 & 0.75000 & -0.304 & 0.066 \\
Ge(1)   & $8g$ & 0.890 & 0.800 & 0.34708 & 0.62468 & 0.75000 & -0.291 & -0.012 \\
Ge(1)   & $8g$ & 0.110 & 0.200 & 0.84708 & 0.62468 & 0.75000 & 0.347 & -0.010 \\
Ge(2)   & $4c$ & 1.000 & 1.000 & 0.00000 & 0.54201 & 0.25000 & -0.007 & 0.023 \\
Ge(3)   & $4c$ & 1.000 & 1.000 & 0.50000 & 0.70488 & 0.25000 & -0.005 & 0.027 \\
Ge(4)   & $4c$ & 1.000 & 1.000 & 0.50000 & 0.78947 & 0.75000 & -0.007 & 0.025 \\
Ge(5)   & $4c$ & 1.000 & 1.000 & 0.50000 & 0.45961 & 0.75000 & -0.006 & 0.031 \\
Fe(1)   & $8e$ & 1.000 & 1.000 & 0.24961 & 0.50000 & 0.50000 & -0.010 & 0.020 \\
Fe(2)   & $8g$ & 1.000 & 1.000 & 0.25046 & 0.62494 & 0.25000 & -0.011 & 0.013 \\
Fe(3)   & $8d$ & 1.000 & 1.000 & 0.75000 & 0.75000 & 0.50000 & 0.000 & 0.020 \\
\bottomrule
\end{tabular}
\end{table*}

\begin{figure*}[htbp] 
    \centering
    \includegraphics[width=1.0\textwidth]{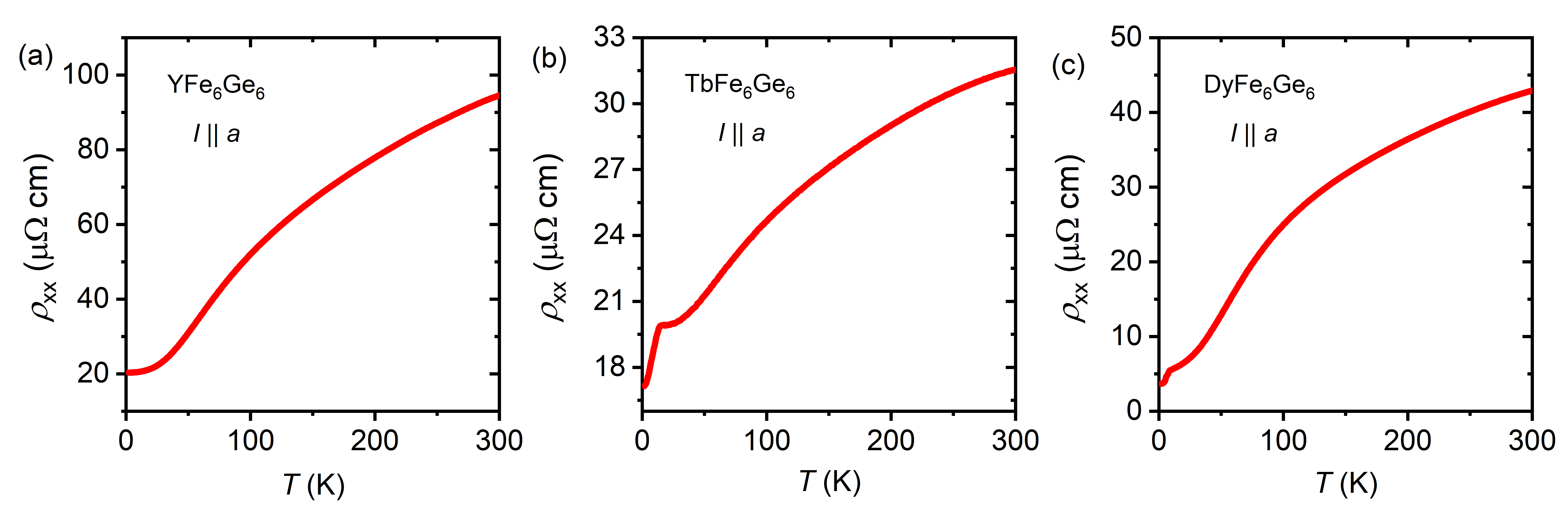}
    \caption{\small Temperature-dependent longitudinal resistivity measurements were reproduced using a separate batch of single-crystalline samples for (a) YFe$_6$Ge$_6$, (b) TbFe$_6$Ge$_6$, and (c) DyFe$_6$Ge$_6$. The measurements were performed following identical experimental protocols to ensure consistency and to validate the reproducibility of the transport behavior.}
    \label{RT REPRODUCED}
\end{figure*}

\begin{figure*}[htbp] 
    \centering
    \includegraphics[width=1.0\textwidth]{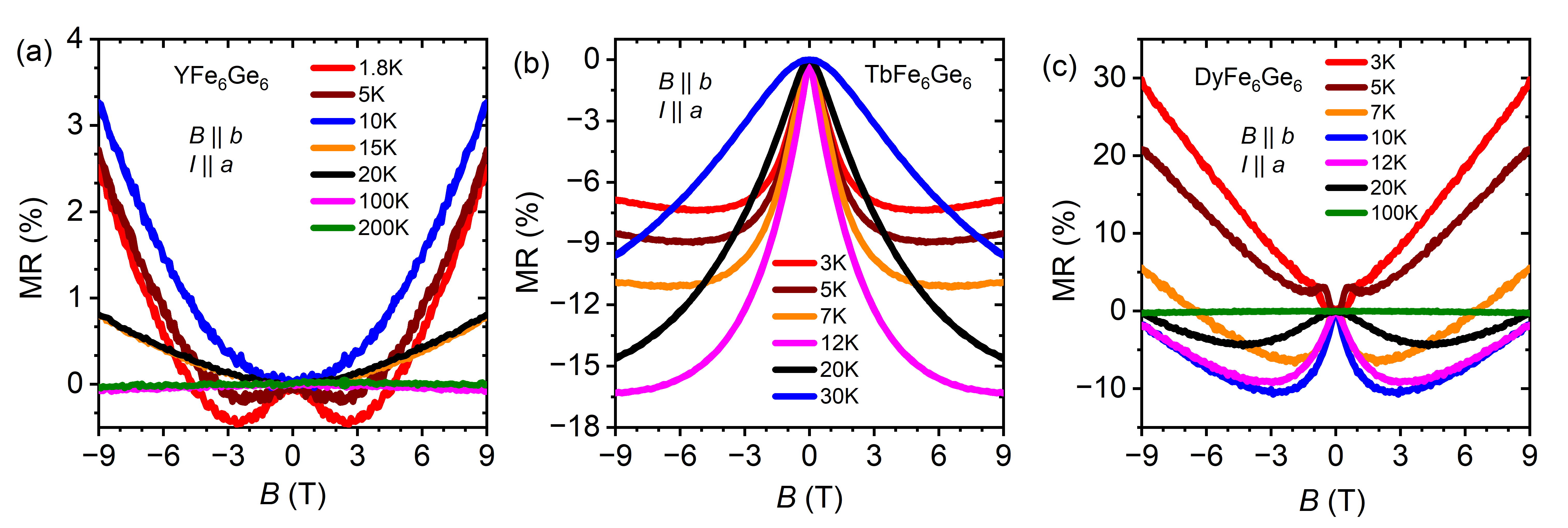}
    \caption{\small Reproducibility of transverse magnetoresistance (MR) measurements was verified using a separate batch of samples for (a) YFe$_6$Ge$_6$, (b) DyFe$_6$Ge$_6$, and (c) TbFe$_6$Ge$_6$. The measurements were conducted following the same experimental configuration and protocols to ensure consistency across different sample batches.}
    \label{MR REPRODUCED}
\end{figure*}

\begin{figure*}[htbp] 
    \centering
    \includegraphics[width=1.0\textwidth]{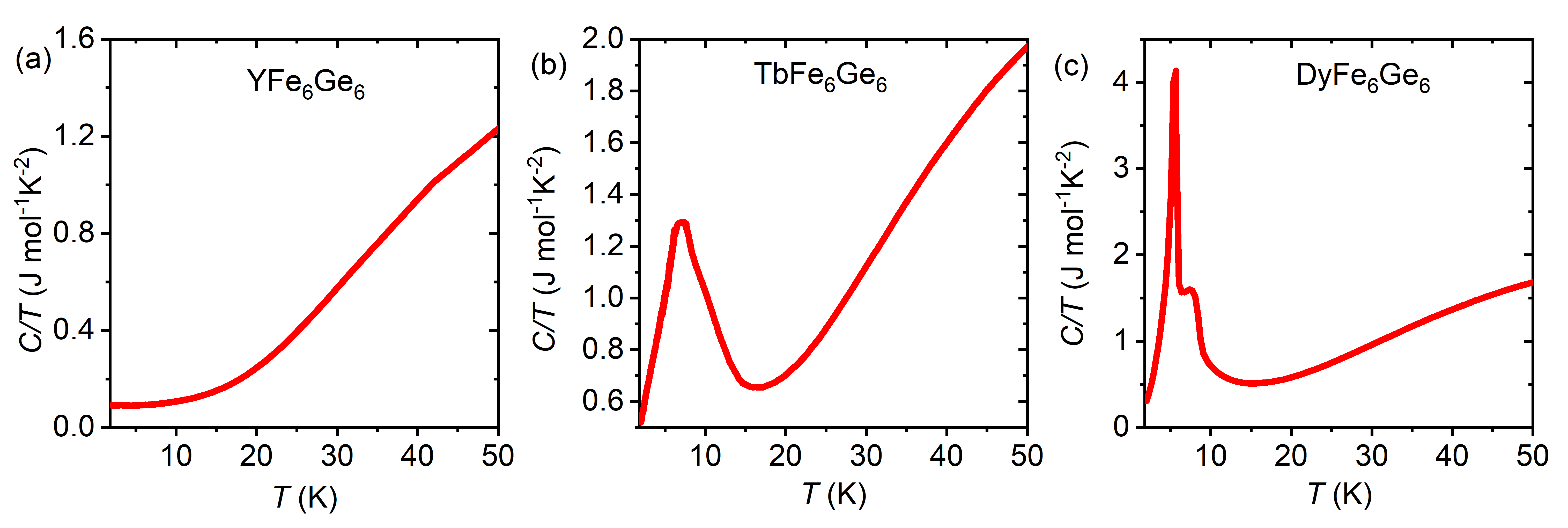}
    \caption{\small Reproducibility of specific heat measurements, plotted as $C/T$ versus temperature, was confirmed using a separate batch of samples for (a) YFe$_6$Ge$_6$, (b) TbFe$_6$Ge$_6$, and (c) DyFe$_6$Ge$_6$. All measurements were performed under identical conditions to ensure consistency and to validate the intrinsic thermodynamic behavior of each compound.}
    \label{HC REPRODUCED}
\end{figure*}

\begin{figure*}[htbp] 
    \centering
    \includegraphics[width=1.0\textwidth]{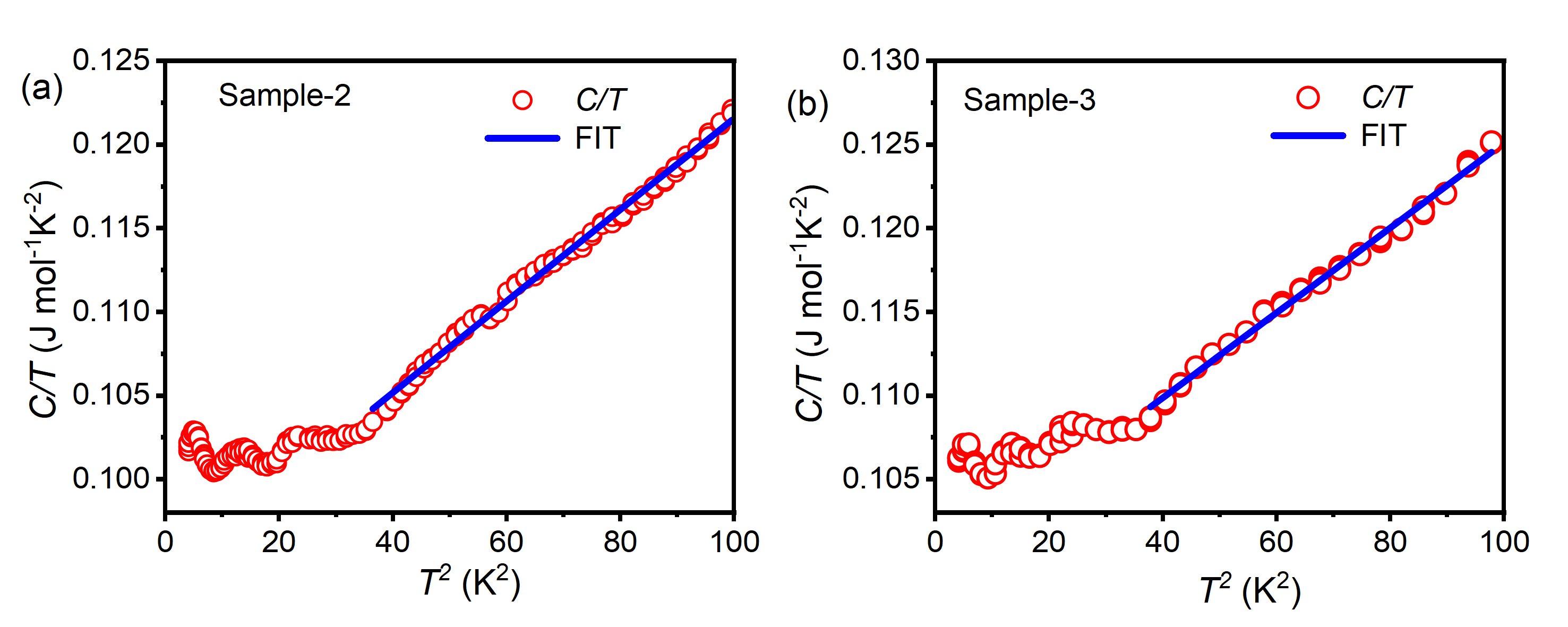}
    \caption{\small Reverification of specific heat measurements from different batch of YFe$_6$Ge$_6$ from 1.8K to 10K and plotted as $C/T$ versus $T$$^2$(a) sample 2 from batch 2, (b) sample 3 from batch 3.}
    \label{Y166 HC Reproduced}
\end{figure*}

\end{document}